\documentclass{jpp}

\usepackage{graphicx}
\usepackage{natbib}
\usepackage{amsmath}
\usepackage{color}

\ifCUPmtlplainloaded \else
  \checkfont{eurm10}
  \iffontfound
    \IfFileExists{upmath.sty}
      {\typeout{^^JFound AMS Euler Roman fonts on the system,
                   using the 'upmath' package.^^J}%
       \usepackage{upmath}}
      {\typeout{^^JFound AMS Euler Roman fonts on the system, but you
                   dont seem to have the}%
       \typeout{'upmath' package installed. JPP.cls can take advantage
                 of these fonts, if you use 'upmath' package.^^J}%
      }
  \else
  \fi
\fi

\ifCUPmtlplainloaded \else
  \checkfont{msam10}
  \iffontfound
    \IfFileExists{amssymb.sty}
      {\typeout{^^JFound AMS Symbol fonts on the system, using the
                'amssymb' package.^^J}%
       \usepackage{amssymb}%

      }{}
  \fi
\fi

\ifCUPmtlplainloaded \else
  \IfFileExists{amsbsy.sty}
    {\typeout{^^JFound the 'amsbsy' package on the system, using it.^^J}%
     \usepackage{amsbsy}}
    {}
\fi

\newsavebox{\astrutbox}
\sbox{\astrutbox}{\rule[-5pt]{0pt}{20pt}}

\newcommand{\Alfven}{Alfv\'{e}n }
\newcommand{\Alfvenic}{Alfv\'{e}nic }

\newcommand{\V}[1]{\mathbf{#1}} 

\title[Field-Particle Correlations in Gyrokinetic Turbulence]
      {Diagnosing collisionless energy transfer using field-particle
        correlations: gyrokinetic turbulence}

\author[Klein, Howes, and TenBarge]%
{K\ls R\ls I\ls S\ls T\ls O\ls
P\ls H\ls E\ls R\ns G.\ns K\ls L\ls E\ls I\ls N%
\thanks{Email address
for correspondence: kriskl@umich.edu}$^{1}$,
G\ls R\ls E\ls G\ls O\ls R\ls Y\ns G.\ns H\ls O\ls W\ls E\ls S$^2$
\and J\ls A\ls S\ls O\ls N\ns M.\ns T\ls E\ls N\ls B\ls A\ls R\ls G\ls E$^3$}

\affiliation{$^1$CLASP,
University of Michigan, Ann Arbor MI 48109, USA \\
[\affilskip]
$^2$Department of Physics and Astronomy, 
University of Iowa,
Iowa City, IA 52242, USA\\
[\affilskip]
$^3$IREAP
University of Maryland,
College Park, MD 20742, USA}

\pubyear{2016}
\volume{?}
\pagerange{?}
\date{?; revised ?; accepted ?.}

\begin{document}

\maketitle

\begin{abstract}
Determining the physical mechanisms that extract energy from turbulent
fluctuations in weakly collisional magnetized plasmas is necessary for
a more complete characterization of the behavior of a variety of space
and astrophysical plasmas. Such a determination is complicated by the
complex nature of the turbulence as well as observational constraints,
chiefly that \emph{in situ} measurements of such plasmas are typically
only available at a single point in space. Recent work has shown that
correlations between electric fields and particle velocity
distributions constructed from single-point measurements produce a
velocity-dependent signature of the collisionless damping
mechanism. We extend this work by constructing field-particle
correlations using data sets drawn from single points in strongly
driven, turbulent, electromagnetic gyrokinetic simulations to
demonstrate that this technique can identify the collisionless
mechanisms operating in such systems. The correlation's velocity-space
structure agrees with expectations of resonant mechanisms transferring
energy collisionlessly in turbulent systems.  This work motivates the
eventual application of field-particle correlations to spacecraft
measurements in the solar wind, with the ultimate goal to determine
the physical mechanisms that dissipate magnetized plasma turbulence.
\end{abstract}

\section{Introduction}
\label{sec:intro}

Studies of the turbulent transport of mass, momentum, and energy have
composed a significant fraction of plasma physics research over the
last half century. Of particular interest is the question of what
mechanisms extract energy from the turbulent
cascade, damping electromagnetic fluctuations and eventually
irreversibly heating the plasma. The answer to this question will
improve the understanding of a wide array of plasma systems, ranging
from laboratory devices, planetary magnetospheres, the Sun and its
extended atmosphere, and accretion disks around massive astrophysical
bodies.

Proposed mechanisms for the dissipation of turbulence in weakly
collisional plasmas can broadly be grouped into three classes:
resonant mechanisms, such as Landau damping, transit-time damping, or
cyclotron damping
\citep{Landau:1946,Barnes:1966,Coleman:1968,Denskat:1983,Isenberg:1983,Goldstein:1994,Quataert:1998,Leamon:1998b,Gary:1999,Hollweg:2002,TenBarge:2013a};
nonresonant mechanisms, such as the stochastic heating of ions in large
amplitude, low-frequency \Alfvenic turbulence
\citep{McChesney:1987,Chen:2001,Johnson:2001,Bourouaine:2008,Chandran:2010a,Chandran:2010b,Bourouaine:2013};
and intermittent dissipation in current sheets and magnetic
reconnection sites
\citep{Dmitruk:2004,Markovskii:2011,Matthaeus:2011,Servidio:2011a,Karimabadi:2013,Zhdankin:2013,Osman:2014a,Osman:2014b,Zhdankin:2015a}.
All three classes couple electromagnetic fluctuations to the plasma
particle velocity distribution, leading to energy transfer and
damping. This coupling occurs through the nonlinear wave-particle
interaction term in the Vlasov equation
\citep{Howes:2015b,Howes:2017a}. Each mechanism produces distinct
structures in velocity space that are characteristic of that
mechanism. For example, resonant mechanisms preferentially energize
particles near a resonant velocity with a change in sign in the energy
transfer across that resonant velocity, while the stochastic heating
described by \citet{Chandran:2010a} will only energize thermal and
sub-thermal particles, producing a platykurtic distribution
\citep{Klein:2016a}.

The solar wind, a low-density and high-temperature plasma accelerated
from the Sun that flows radially outward through the heliosphere,
is a heavily sampled space plasma system, with measurements dating
back to the dawn of the space age. The large quantity of measurements of
this super-\Alfvenic plasma flow makes it a unique system with which
various plasma and turbulence theories, including descriptions of
damping and dissipation, can be tested across a wide range of physical
scales and plasma parameters. A significant limitation of such \emph{in situ}
measurements is that most observations occur at a single point, or at
most a few points, in space. Any attempts to use the solar wind to
test theories of turbulence must take this limitation into
consideration.


A novel field-particle correlation technique has been proposed
\citep{Klein:2016,Howes:2017a} which uses single-point measurements to
capture the non-oscillatory, or secular, transfer of energy associated
with the net removal of energy from turbulent fluctuations. This
correlation isolates the secular energy transfer by averaging the
nonlinear field-particle interaction term in the Vlasov equation over
a time interval longer than the time scale characteristic of the
turbulent fluctuations involved in the energy transfer. Crucially, not
only can the secular energy transfer be isolated, but the
velocity-space structure of this correlation can be used to
discriminate between various collisionless damping mechanisms using
only single-point measurements of the type accessible to spacecraft in
the solar wind.

A fundamental question at the forefront of heliophysics research is
whether Landau damping, and other resonant wave-particle interactions,
can effectively remove energy from turbulent plasmas given the highly
nonlinear nature of strong plasma turbulence
\citep{Plunk:2013,Schekochihin:2016}. In this work, we seek evidence
of Landau damping in turbulent plasmas by the application of
field-particle correlations to data extracted from a series of
gyrokinetic simulations of 3D electromagnetic plasma turbulence. This
extends previous work applying such correlations to monochromatic,
electrostatic waves \citep{Klein:2016,Howes:2017a,Klein:2017} and the
Landau damping of a single kinetic \Alfven wave \citep{Howes:2017b} to
a system of strong turbulence where the role of resonant interactions
is a matter of current debate.

The remainder of this paper is organized as follows.  The
electromagnetic form of the field-particle correlation is developed in
Sec.~\ref{sec:fpem}. A presentation of the expected structure of
damping in low-frequency turbulence is found in Sec.~\ref{sec:mech},
followed by a discussion of the simulation code employed,
\texttt{AstroGK}, in Sec.~\ref{sec:lft}. In Sec.~\ref{sec:KAW}, we
apply the correlation to a single kinetic \Alfven wave, followed by an
application to turbulent simulations in Sec.~\ref{sec:apply}. Discussion,
summary, and future applications are found in Sec.~\ref{sec:concl}.
Even in the presence of strong turbulence and spatially inhomogeneous
heating, the secular energy transfer from the turbulent fields to the
protons is shown to be localized in velocity space near the resonant
velocities associated with Landau damping. This work motivates future
application to data from turbulence simulations which contain other
damping mechanisms, as well as spacecraft observations, with the
ultimate goal to determine which mechanisms act to dissipate
turbulence in the solar wind.

\section{Field-Particle Correlations for Electromagnetic Fluctuations}
\label{sec:fpem}
The novel approach of using field-particle correlations to diagnose
the energy transfer between fields and particles has been described
for electrostatic fluctuations in the Vlasov-Poisson
system \citep{Howes:2017a} and been applied to both damped and linearly
unstable systems \citep{Klein:2016,Klein:2017,Howes:2017b}.  Here we describe the
application of the field-particle correlation technique to the case of
electromagnetic fluctuations in the Vlasov-Maxwell system.

The Boltzmann equation describes the dynamics and energetics of weakly
collisional plasmas relevant to heliospheric environments, such as the
solar corona and the solar wind, determining the evolution of the
six-dimensional velocity distribution function $f_s(\V{r},\V{v},t)$
for a plasma species $s$,
\begin{equation}
\frac{\partial f_s}{\partial t} + \mathbf{v}\cdot \nabla f_s + \frac
     {q_s}{m_s}\left[ \mathbf{E}+ \frac{\mathbf{v} \times \mathbf{B}
       }{c} \right] \cdot \frac{\partial f_s}{\partial \mathbf{v}} =
     \left(\frac{\partial f_s}{\partial t} \right)_{\mbox{coll}}.
\label{eq:boltzmann}
\end{equation}
Combining the Boltzmann equation for each plasma species together with
Maxwell's equations forms the closed set of Maxwell-Boltzmann
equations that govern the nonlinear evolution of turbulent
fluctuations in a magnetized kinetic plasma.

Here we focus strictly on the collisionless dynamics of the energy
transfer between fields and particles, so we drop the collision
operator on the right-hand side of \eqref{eq:boltzmann} to obtain the
Vlasov equation.  Multiplying the Vlasov equation by $m_s v^2/2$,
integrating over all position and velocity space, and using an
integration by parts in velocity for the Lorentz force term yields the
expression
\begin{equation}
  \frac{\partial W_s}{\partial t} =  \int d^3\V{r} \int d^3\V{v}
  \ q_s \V{v} \cdot \V{E}  f_{s} = \int d^3\V{r} \ \V{j}_s \cdot \V{E},
  \label{eq:dwsdt}
\end{equation}
where the current density of species $s$ is defined as $\V{j}_s\equiv
\int d^3\V{v} q_s\V{v}f_s$ and the \emph{microscopic particle kinetic
  energy} for species $s$ is given by
\begin{equation}
W_s \equiv \int d^3\V{r} \int d^3\V{v} \ \frac{1}{2}m_s v^2 f_s(\V{r},\V{v},t).
\end{equation}
Note that the ballistic term (the second term on the left-hand side of
\eqref{eq:boltzmann}) and the magnetic part of the Lorentz term (the
third term on the left-hand side) yield zero net energy transfer upon
integration over all position and velocity space, assuming suitable
boundary conditions, such as periodic boundaries or infinitely distant
boundaries with vanishing $f_s$.  Summing \eqref{eq:dwsdt} over
species and combining with Poynting's Theorem
\begin{equation}
  \frac{\partial }{\partial t}  \int d^3\V{r} \frac{|\V{E}|^2+|\V{B}|^2 }{8\pi}
  + \frac {c}{4\pi} \oint  d^2\V{S} \cdot ( \V{E}\times\V{B}) =
  -  \int d^3\V{r}\; \; \V{j} \cdot \V{E},
  \label{eq:poynting}
\end{equation}
we may obtain the  \emph{conserved Vlasov-Maxwell energy}, $W$, for
electromagnetic fluctuations in a collisionless, magnetized plasma,
\begin{equation}
  W =  \int d^3\V{r} \frac{|\V{E}|^2+|\V{B}|^2 }{8\pi}
  + \sum_s  \int d^3\V{r} \int d^3\V{v} \ \frac{1}{2}m_s v^2 f_s.
  \label{eq:vmcons}
\end{equation}
Note that, for periodic or infinitely distant boundaries, the Poynting
flux term, the second term on the left hand side of
\eqref{eq:poynting}, yields zero net change in the energy $W$.

In the Vlasov-Maxwell system, \eqref{eq:dwsdt} shows clearly that the
change in the microscopic energy of the particles is accomplished by
interactions of the particles with the electric field, where $\V{j}_s
\cdot \V{E}$ is the (spatially) local rate of change of the energy
density of particle species $s$. But, as pointed out in the previous
description of how to use field-particle correlations to explore the
conversion of turbulent energy into microscopic particle energy
\citep{Howes:2017a}, this energy transfer between fields and particles
includes both the conservative \emph{oscillating energy transfer}
associated with undamped wave motion and the \emph{secular energy
  transfer} associated with the collisionless damping of the turbulent
fluctuations. Here we specifically define the turbulence as the sum of
the fluctuations in the electromagnetic fields and the fluctuations of
the bulk flows of the plasma \citep{Howes:2015b}.  Collisionless
interactions between the fields and the particles, governed by the
Lorentz force term (third term on the left-hand side) in
\eqref{eq:boltzmann}, remove the energy from the turbulent
fluctuations, transferring it into microscopic particle kinetic energy
that is not associated with bulk plasma motions.  Diagnosing this net
transfer of energy between fields and particles is the key aim of the
field-particle correlation method, using a time-average over an
appropriately chosen correlation interval to eliminate the often large
signal of the oscillating energy transfer, exposing the smaller signal
of the secular energy transfer.

A significant limitation of spacecraft measurements is that
information is generally limited to a single point, or at most a few
points, in space.  Therefore, the spatial integration necessary to
simplify the energy transfer in the Vlasov equation to the form given
by \eqref{eq:dwsdt} is not possible. To explore the energy transfer
between fields and particles at a single point in space, we define the
\emph{phase-space energy density} for a particle species $s$ by
$w_s(\V{r},\V{v},t) = m_s v^2 f_s(\V{r},\V{v},t)/2$. Multiplying the
Vlasov equation by $ m_s v^2/2$, but not integrating over space or
velocity, we obtain an expression for the rate of change of the
phase-space energy density,
\begin{equation}
  \frac{\partial w_s(\V{r},\V{v},t)}{\partial t} = - \V{v}\cdot \nabla  w_s  -
  q_s\frac{v^2}{2}  \mathbf{E} \cdot \frac{\partial f_s}{\partial \mathbf{v}}
   - \frac{q_s}{c}\frac{v^2}{2} \left(\mathbf{v} \times \mathbf{B}\right)
      \cdot \frac{\partial f_s}{\partial \mathbf{v}}.
  \label{eq:dws}
\end{equation}
When integrated over velocity space, an integration by parts of the
last term on the right-hand side of \eqref{eq:dws} yields an integrand
containing $\mathbf{v} \cdot \left(\mathbf{v} \times
\mathbf{B}\right)=0$, so the magnetic field cannot accomplish any net
change of energy of the particles. In addition, when integrated over
volume, the first term on the right-hand side of \eqref{eq:dws} yields
zero net energy change for either periodic or infinite boundary
conditions.  Therefore, we focus here on the second term on the
right-hand side of \eqref{eq:dws}, the term that determines the effect
of the electric field on the rate of change of phase-space energy
density.\footnote{As discussed in \cite{Howes:2017a}, it is imperative
  to use a consistent frame of reference for the particle distributions
  and the fields, which in the case of spacecraft measurements will
  require a Lorentz transformation of the electric field from the
  spacecraft to the plasma frame.}

Examining the electric-field term, we can write the form of the
field-particle correlation at a single point $\V{r}_0$ for the general
Vlasov-Maxwell case, separating the contributions from the parallel
and perpendicular parts of the electric field,
\begin{equation}
  C_{E_\parallel} (\V{v},t,\tau)= C\left(-  q_s\frac{v_\parallel^2}{2} \frac{\partial f_s(\V{r}_0,\V{v},t)}{\partial v_\parallel},E_\parallel(\V{r}_0,t)\right)
   \label{eq:cepar}
\end{equation}
\begin{equation}
  C_{E_\perp}(\V{v},t,\tau) = C\left(-  q_s\frac{v_x^2}{2} \frac{\partial f_s(\V{r}_0,\V{v},t)}{\partial {v}_x},{E}_x(\V{r}_0,t)\right)
  +  C\left(-  q_s\frac{v_y^2}{2} \frac{\partial f_s(\V{r}_0,\V{v},t)}{\partial {v}_y},{E}_y(\V{r}_0,t)\right),
   \label{eq:ceperp}
\end{equation}
where we define the unnormalized correlation
\begin{equation}
C(A,B)\equiv \frac{1}{N}\sum_{j=i}^{i+N}A_jB_j
\label{eqn:corr}
\end{equation}
for quantities $A$ and $B$ measured at discrete times $t_j = j \Delta t$
with correlation interval $\tau \equiv N \Delta t$.
Note that the $v^2=v_\parallel^2+v_\perp^2$ factor is reduced to
$v_\parallel^2$ for $ C_{E_\parallel}$ because the net energy change
is zero for the $v_\perp^2$ contribution when integrated over
velocity. Similarly, for $ C_{E_\perp}$, one uses $v_x^2$ for the
energy change due to $E_x$ and $v_y^2$ for the energy change due to
$E_y$, where we assume for notational simplicity that the local
magnetic field is in the $\hat{\V{z}}$ direction, $\V{B} = B(r_0)\hat{\V{z}}$.

Note that, depending on the physical mechanism to be investigated, one
will choose the appropriate correlation, either $C_{E_\parallel}$ or
$C_{E_\perp}$. One would choose $C_{E_\parallel}$ to investigate
Landau damping, since it is mediated by the parallel electric field,
and one would choose $C_{E_\perp}$ to study cyclotron damping or
stochastic ion heating, since these mechanisms are mediated by the
perpendicular electric field. Importantly, regardless of the
underlying mechnanism, $C_{E_\parallel}$ and $C_{E_\perp}$ measure the
energy density transfer mediated by the associated electric field
component.

Since taking the velocity gradient of noisy or low resolution
phase-space measurements of particle velocity distribution functions
can lead to large errors, one can define a related correlation $C'$
that is derived by an integration by parts in velocity \citep{Howes:2017a}.  These
alternative forms are
\begin{equation}
  C'_{E_\parallel} (\V{v},t,\tau)= C\left(  q_s v_\parallel f_s(\V{r}_0,\V{v},t),E_\parallel(\V{r}_0,t)\right)
   \label{eq:cprime_epar}
\end{equation}
\begin{equation}
  C'_{E_\perp}(\V{v},t,\tau) = C\left(  q_sv_x f_s(\V{r}_0,\V{v},t),{E}_x(\V{r}_0,t)\right)
  +  C\left(  q_s v_y f_s(\V{r}_0,\V{v},t),{E}_y(\V{r}_0,t)\right).
   \label{eq:cprime_eperp}
\end{equation}
When integrated over velocity, $C'_{E_\parallel}$ simply yields the
time-averaged $j_{s \parallel} E_\parallel$ and $C'_{E_\perp}$ yields
the time-averaged $\V{j}_{s\perp}\cdot \V{E}_\perp$, which is the net
electromagnetic work done on the particles.  And since the alternative
forms $C'$ are equivalent to the $C$ forms of the field-particle
correlations when integrated over velocity, the original forms $C$
given by \eqref{eq:cepar} and \eqref{eq:ceperp} also yield, upon
velocity integration, the net electromagnetic work done on the
particles.

\section{Resonant Energy Transfer in Low-Frequency Turbulence}
\label{sec:mech}

We next describe the damping mechanisms accessible to turbulence in
solar and astrophysical plasmas and predict where in velocity space
the related energy transfer is expected to appear.  The resonant
mechanisms that can remove energy from weakly collisional plasma
turbulence depend strongly on the frequency of the turbulent
fluctuations.  Direct multi-spacecraft observations of the solar wind
find that turbulent fluctuations at length scales near ion kinetic
scales are anisotropic, with $k_\perp \gg k_\parallel$, with $\perp$
and $\parallel$ defined with respect to the local mean magnetic field
\citep{Sahraoui:2010b,Narita:2011,Roberts:2013,Roberts:2015b}.  The
frequency of such fluctuations depends on the normal-mode response of
the plasma: fast/whistler fluctuations have frequencies $\omega \sim
\Omega_p$ at $k_\perp d_p \sim 1$, while both Alfv\'en/kinetic \Alfven
waves (KAWs) and slow/kinetic slow waves have lower frequencies,
$\omega \ll \Omega_p$, where $\Omega_p$ is the proton cyclotron
frequency and $d_p=v_A/\Omega_p$ is the proton inertial length
\citep{Howes:2012a,Howes:2014a}.  Theoretical
\citep{Schekochihin:2009}, numerical \citep{Howes:2008a}, and
observational \citep{Sahraoui:2010b} studies of solar and
astrophysical plasmas find that $\omega \ll \Omega_p$, suggesting that
fast/whistler fluctuations play a minor role in governing the
turbulence of these systems.  Both historic and recent \emph{in situ}
observations of the solar wind find that the measured fluctuations
have polarizations consistent with low-frequency \Alfven waves at
larger scales \citep{Belcher:1971} and KAWs at ion scales
\citep{Salem:2012,Podesta:2012,TenBarge:2012b,Chen:2013a,Kiyani:2013}.
This body of evidence suggests that the turbulence is dominated by
low-frequency, anisotropic \Alfvenic fluctuations with $\omega \ll
\Omega_p$.

Kinetic plasma theory dictates that resonant collisionless damping
mechanisms satisfy the resonance condition $\omega - k_\parallel
v_\parallel -n \Omega_p=0$, where $n$ is any integer.  For
low-frequency, anisotropic turbulence with $\omega \ll \Omega_p$, the
collisionless damping arising through the cyclotron, or $n \neq 0$,
resonances is expected to be negligible \citep{Lehe:2009}.  The
dominant resonant damping mechanisms in such systems are those which
approximately satisfy the $n=0$ condition $\omega - k_\parallel
v_\parallel =0$, known as the Landau resonance. Two specific
collisionless damping mechanisms, Landau damping \citep{Landau:1946}
and transit time damping \citep{Barnes:1966}, operate via this
resonance; the former is mediated by the electric force of the
parallel electric field on the particle charge, while the latter is
mediated by the magnetic mirror force from the variations in the
magnetic field magnitude on the particle magnetic moment $\mu_s = m_s
v_\perp^2/2|B|$. In this work, we focus on capturing the signature of
Landau damping using the field-particle correlation $C_{E_\parallel}$
in strong plasma turbulence. Application of the perpendicular electric
field correlation $C_{E_\perp}$, which is expected to capture
cyclotron damping, will be considered in later work using a simulation
model that captures the physics of the cyclotron resonance, since the
gyrokinetic approximation orders out the cyclotron physics,
eliminating this damping mechanism in gyrokinetic simulations
\citep{Howes:2006}.

Previous applications of field-particle correlations of the type
described in Section~\ref{sec:fpem} were used on relatively simple
systems of one or a few wave modes which damp via resonant
interactions with one velocity
\citep{Klein:2016,Howes:2017a,Howes:2017b} or a few velocities
\citep{Klein:2017}.  One might naively expect a spectrum of turbulent
fluctuations potentially to interact with a broad range of velocities,
eliminating any coherent signature that may be used to diagnose the
nature of the damping mechanism. However, by examining the linear
properties of low-frequency \Alfven waves, we predict that the
turbulent fluctuations should preferentially interact with ions over a
relatively narrow band of resonant velocities.

\begin{figure}
  \centerline{\includegraphics[width = 18cm]
{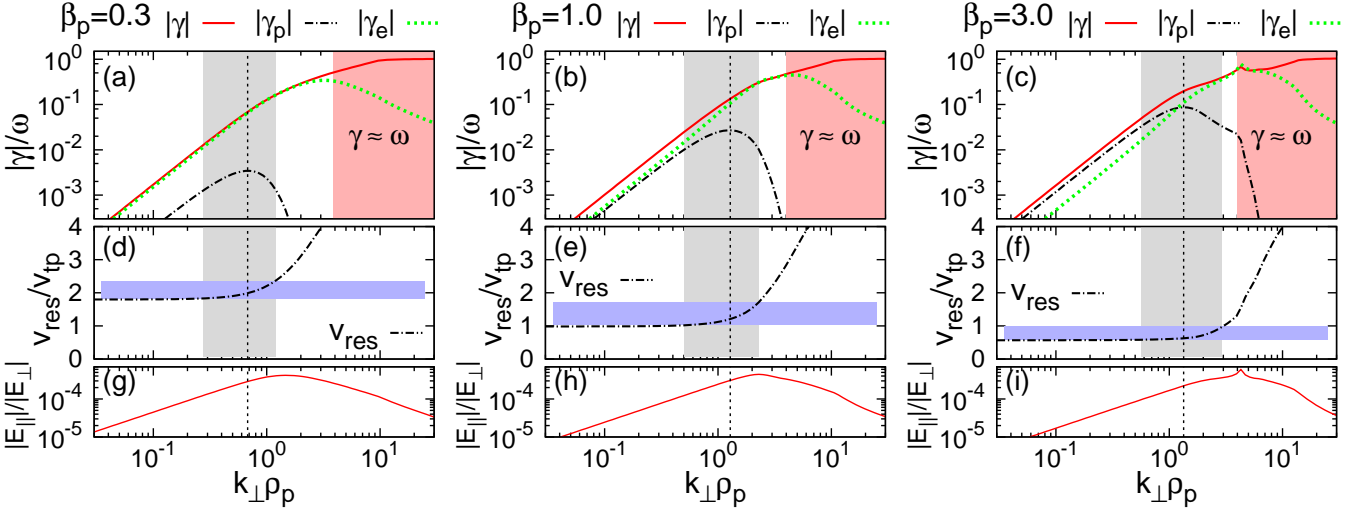}}%
  \caption{Linear characteristics of the collisionless, low-frequency
    \Alfven dispersion relation as a function of $k_\perp \rho_p$ for
    $\beta_p=0.3$ (left row), $1.0$ (center), and $3.0$ (right) and
    reduced mass ratio $m_p/m_e=32$. Panels (a-c) plot the linear
    damping rate $|\gamma|/\omega$ (red line) and proton
    $|\gamma_p|/\omega$ (black) and electron $|\gamma_e|/\omega$ (green)
    power absorption.  The resonant proton velocities (black lines)
    are plotted in panels (d-f) and the electric field ratio
    $|E_\parallel|/|E_\perp|$ (red lines) is shown in panels (g-i).}
\label{fig:linear}
\end{figure}

In Fig.~\ref{fig:linear}, linear Vlasov-Maxwell damping rates for an
\Alfven wave are presented as a function of $k_\perp \rho_p$ with
constant $k_\parallel \rho_p = 10^{-3}$, where $\rho_p$ is the proton
gyroradius. The following plasma parameters were used: $T_p=T_e$,
$T_{\perp,s}=T_{\parallel,s}$, $v_{tp}/c=10^{-4}$, $m_p/m_e=32$, with
$\beta_p=0.3,1.0,$ and $3.0$, where $\beta_p=8\pi n_p T_p/B^2$.  All
of the characteristics of the linear, magnetized, collisionless,
fully-ionized, proton-electron plasma response are calculated using
the the \texttt{PLUME} dispersion solver \citep{Klein:2015a}.  We
constrain the values of $\beta_p$ to those near unity to match typical
solar wind observations. Additionally, we model a plasma with a
reduced mass ratio, shifting significant electron damping to larger
scales by the reduction of the ratio of the species Larmour radii,
$\rho_e/\rho_p=\sqrt{m_e/m_p}$.\footnote{We have ensured that our
  selected mass ratio is sufficiently large that the proton damping
  rate decreases at scales $k_\perp \rho_p \gg 1$ in a quantitatively
  similar fashion to the proton damping rate for a plasma with a
  realistic mass ratio of $m_p/m_e = 1836$.}  With this choice of
reduced mass ratio, there is sufficient resolved collisionless damping
by the electrons within the limited dynamic range of our simulations
to achieve a steady-state turbulent cascade, with no need for
artificial dissipation at small scales, which could corrupt our
results.

The collisionless power absorption by species $s$ due to a normal mode
in one wave period is calculated following \cite{Stix:1992} $\S 11.8$
and \cite{Quataert:1998} as
\begin{equation}
\frac{\gamma_s}{\omega} = \frac{\V{E}^*\cdot
  \underline{\underline{\chi}}_s^a \cdot \V{E}}{4 W_{\rm EM}},
\label{eqn:power}
\end{equation}
where $\underline{\underline{\chi}}_s^a$ is the anti-Hermitian part of
the linear susceptibility tensor for species $s$ evaluated at the real
component of the normal-mode frequency, $\V{E}$ and $\V{E}^*$ are the
vector electric field associated with the normal mode and its complex
conjugate, and $W_{\rm EM}$ is the electromagnetic wave energy. The total
damping rate is the sum $\gamma=\gamma_p + \gamma_e$. Values for
$\gamma_p$ and $\gamma_e$ are shown in panels (a-c) of
Fig.~\ref{fig:linear}.  As noted in the cited literature, the calculation
of $\gamma_s$ breaks down for $\gamma/\omega \gtrsim 1$; the region
for which $\gamma/\omega \gtrsim 1$ is highlighted in red.

While the total damping rate monotonically increases near $k_\perp
\rho_p = 1.0$, the power absorbed by the protons is a strongly peaked
function near that scale. Proton damping will be dominated by modes
with wavevectors near this peak, which have resonant velocities
bounded within a narrow region.  This peak arises because Landau
damping efficiently operates when both the resonant velocity $v_{\rm
  res}=\omega/k_\parallel$ lies in the bulk of the velocity
distribution and there exists a finite parallel electric field. The
resonant velocity of an \Alfven wave normalized by the thermal
velocity $v_{tp}$, plotted in panels (d-f) of Fig.~\ref{fig:linear},
is shown to be non-dispersive and in resonance with the bulk of the
proton velocity distribution until it reaches length scales of order
$\rho_p$ where the wave transitions to a kinetic \Alfven wave. For the
parameters under consideration, the dispersive modifications increase
$\omega$, moving the wave largely out of resonance with the protons
and reducing damping at small scales. \Alfven waves with $k_\perp
\rho_p \ll 1$ have weak parallel electric fields, as shown in panels
(g-i), which limit the effectiveness of resonant damping at large
scales. Therefore, the proton power absorption in a wave period
$\gamma_p/\omega$ peaks near $k_\perp \rho_p=1$.  The wavevector
region having proton power absorption within one e-folding of the
maximum proton power absorption is highlighted in vertical grey bands
in the top two rows of Fig.~\ref{fig:linear}. The secular energy
transfer to the protons from a spectrum of low-frequency, \Alfvenic
fluctuations should therefore be constrained to a narrow band of
parallel velocities (horizontal blue bands in second row) near the
proton thermal velocity in panels (d-f), with a clear dependence on
$\beta_p$.


\section{Gyrokinetic Simulations of Low-Frequency Turbulence}
\label{sec:lft}

We next detail the turbulence simulations carried out in this study.
Gyrokinetics, a rigorous limit of the Vlasov-Maxwell system of
equations, has been shown to optimally describe the low-frequency,
anisotropic turbulent fluctuations typically found in the solar wind
\citep{Frieman:1982,Howes:2006,Schekochihin:2009}. By averaging over
the gyromotion of the particles, gyrokinetics reduces the
dimensionality of the kinetic system from six (3D-3V) to five
(3D-2V). The gyrokinetic formalism describes damping via the Landau
($n=0$) resonance, \citep{TenBarge:2013a} and resolves the kinetic
microphysics of collisionless magnetic reconnection in the
large-guide-field limit \citep{TenBarge:2014b,Numata:2015}. Mechanisms
such as cyclotron damping and stochastic heating due to low-frequency
\Alfvenic turbulence are not included, the former due to the exclusion
of high-frequency behavior and the latter due to conservation of the
magnetic moment enforced by the gyroaveraging procedure. In this paper,
we focus on recovering the signature of Landau damping, leaving the
identification of other damping mechanisms to later work.

We employ the Astrophysical Gyrokinetics simulation code,
\texttt{AstroGK} \citep{Numata:2010}, which has been used to
successfully model plasma physics phenomena in the heliosphere over
the last decade
\citep{Howes:2008a,Howes:2011a,TenBarge:2012a,TenBarge:2013a,TenBarge:2013c,Numata:2015}.
\texttt{AstroGK} evolves the gyroaveraged scalar potential
$\phi(\V{r})$, parallel vector potential $A_z(\V{r})$, and the
parallel magnetic field fluctuation $\delta B_z(\V{r})$, as well as
the gyrokinetic distribution function
$h_s(\V{R}_s,v_\perp,v_\parallel)$, in a triply-periodic slab geometry.
The gyrokinetic distribution function is related to the total
distribution function $f_s$ via
\begin{equation}
f_s(\V{r}, \V{v}, t) = 
F_{0s}(v)\left( 1 - \frac{q_s \phi(\V{r},t)}{T_{0s}} \right)
+ {h_s}(\V{R}_s, v_\perp, v_\parallel, t) + \delta f_{2s} + ...
\label{eqn:fullF}
\end{equation}
where $F_{0s}$ is the Maxwellian equilibrium distribution, $\V{r}$ is
the spatial position, $\V{R}_s$ the associated species gyrocenter, and
$\delta f_{2s}$ are second-order corrections in the gyrokinetic
expansion parameter $\epsilon \sim k_\parallel/k_\perp$ which are not
retained \citep{Howes:2006}. The domain is a periodic box of size
$L_{\perp }^2 \times L_{\parallel }$, elongated along the straight,
uniform mean magnetic field $\V{B}_0=B_0 \hat{z}$. The code employs a
pseudospectral method in the x-y (perpendicular) plane and
finite-differencing in the z-direction. The velocity distribution is
resolved on a grid in energy $E=v^2/2$ and pitch angle $\lambda =
v_\perp^2/v^2$ space, with the points selected on a Legendre
polynomial basis. A fully conservative, linearized, gyroaveraged
collision operator is employed \citep{Abel:2008,Barnes:2009}.

As a technical step, we transform from the gyrokinetic distribution
function $h_s$ to the complementary perturbed distribution
\begin{equation}
{g_s}(\V{R}_s,v_\perp, v_\parallel) = {h_s}(\V{R}_s,v_\perp, v_\parallel) 
- \frac{q_s F_{0s}}{T_{0s}} 
\left< 
\phi 
- \frac{\V{v}_\perp \cdot \V{A}_\perp}{c}
\right>_{\V{R}_s}
\label{eqn:g+h}
\end{equation}
\citep{Schekochihin:2009}, where $\left < ...\right >$ is the
gyroaveraging operator. The complementary distribution function $g_s$
describes perturbations to the background distribution in the frame
moving with an \Alfven wave.  Such perturbations are associated with
the compressive components of turbulence and therefore are associated
with the collisionless damping mechanism under consideration.
Field-particle correlations calculated using $h_s$ or $f_s$ (not
shown) yield qualitatively and quantitatively similar results to those
computed with $g_s$.

\begin{figure}
  \centerline{\includegraphics[width = 12cm]
{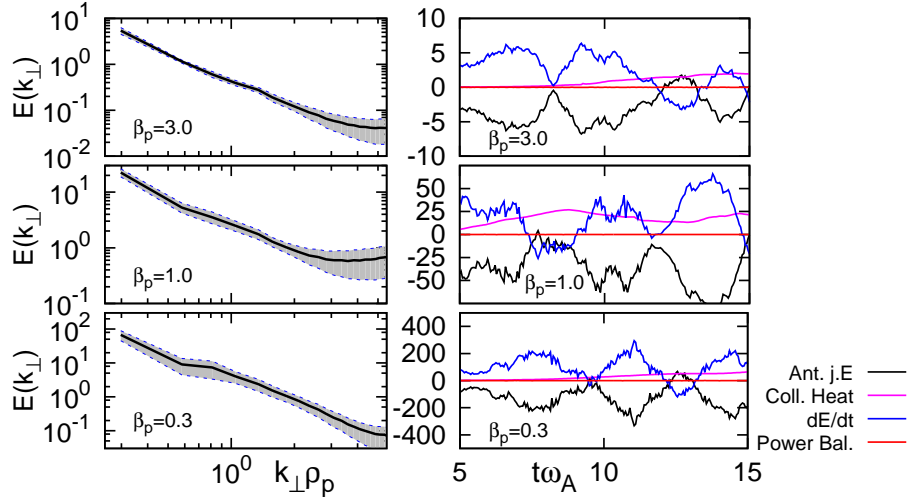}}%
  \caption{Power spectra for the $\beta_p = 0.3, 1.0$ and
    $3.0$ simulations (left column), with the standard deviation of
    the spectra shown as grey shading. An evaluation of the energy
    injected into and collisionally dissipated from the simulations
    (right column) demonstrates the steady-state nature of the
    turbulence.}
\label{fig:sim}
\end{figure}

We perform three turbulent simulations with nearly identical setups.
The number of simulated grid points is $(n_x,n_y,n_z,n_\lambda,n_E)=
(64,64,32,64,32)$, where $n_\lambda$ and $n_E$ are the number of pitch
angle and energy points. The fully resolved simulation domain spans
$k_\perp \rho_p \in [0.25,5.5]$, or $k_\perp \rho_e \in [0.04,0.97]$
for the reduced mass ratio under consideration, $m_p/m_e=32$.  The
maximum $v_\perp$ and $v_\parallel$ resolved for each species is $4
v_{ts}$.  We set $\beta_p$ to $0.3,1.0,$ and $3.0$ for the three
simulations. The simulations are driven using an oscillating Langevin
antenna \citep{TenBarge:2014} that drives fluctuations with
wavevectors $(k_x,k_y,k_z)=(1,0,\pm 1)$ and $(0,1,\pm 1)$ plus their
complex conjugates with amplitudes sufficient to drive the system into
a staturated state of strong turbulence.  All three simulations are
run to at least $t \omega_A = 20$, where $\omega_A = k_\parallel v_A$
with $k_\parallel = 2 \pi / L_\parallel$.  The proton collision
frequency is set at approximately a tenth of the maximum linear proton
damping rate, $\nu_p/k_\parallel v_{tp} = 5\times10^{-5}$, $2 \times
10^{-4}$, and $1\times 10^{-3}$ for the $\beta_p=0.3,1.0,$ and $3.0$
runs respectively.

Power spectra for the three turbulent simulations are shown in
Fig.~\ref{fig:sim}, averaged over an outer-scale \Alfven turn-around
time starting once the turbulence has reached steady state, at around
$t \omega_A = 6$. The surrounding grey shaded regions represent the
standard deviation of the spectra over the time interval used for
averaging. We note there is evidence of bottlenecking (flattening of
the spectra) at the smallest scales in these simulations, as we have
elected to not introduce artificial hypercollisionality which may
obscure signatures of the collisionless damping mechanisms in the
velocity distribution function.  To ensure the simulations are in a
steady state, we evaluate the external energy injected into the system
via the antenna (black lines), the collisional entropy production
(pink), and the time derivative of the fluctuation energy (blue). We
note that their sum (red line) is zero, indicating good conservation
of energy in these simulations.  A detailed discussion of these terms
can be found surrounding equation (B19) in \cite{Howes:2006}.

Both the complementary perturbed distribution
$g_p(\V{r}_j;v_\perp,v_\parallel)$ and $E_\parallel(\V{r}_j)$ are
output at selected fixed points in the spatial domain at a
fixed cadence to mimic single-point observations of the solar
wind. With this single-point diagnostic, we calculate the
field-particle correlation $C_{E_\parallel}$, representing the first
application of this technique to a turbulent data set.

\section{Field-Particle Correlations for a Single KAW}
\label{sec:KAW}

\begin{figure}
  \centerline{\includegraphics[width = 18cm]
    {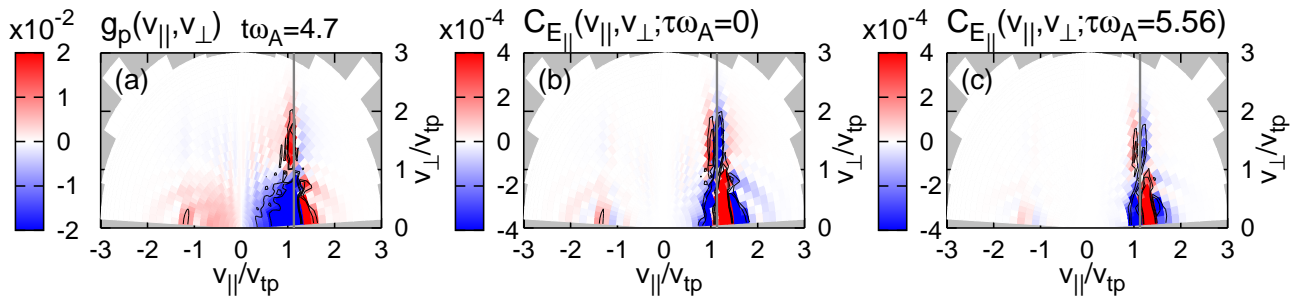}}%
  \caption{The proton gyrotropic complementary distribution function
    at a point in the single KAW simulation, panel (a), and the
    correlations $C_{E_\parallel}(\tau=0)$ and $C_{E_\parallel}(\tau
    \omega_A =5.56)$, panels (b) and (c), at a point in time,
    $t\omega_A=4.7$. The resonant velocity of the KAW is shown as a
    solid grey line.}
\label{fig:gyro_KAW}
\end{figure}

Before applying the field-particle correlations to data from the three
turbulence simulations, we first consider a single, nonlinearly
evolving kinetic \Alfven wave, similar to the case presented in
\cite{Howes:2017b}. We initialize a single KAW with $k_\perp \rho_p =
1$ and $\beta_p = 1.0$, following the eigenfunction initialization
specified in \cite{Nielson:2010}. The gyrotropic complementary proton
distribution at a single point $\V{r}_0$ in the simulation is plotted in
Fig.~\ref{fig:gyro_KAW} at time $t\omega_A = 4.7$. Also plotted are
the instantaneous rate of change of the phase-space energy density,
$C_{E_\parallel}(\tau=0)$, and the time averaged correlation
$C_{E_\parallel}(\tau\omega_A = 5.56)$.  The correlation interval
$\tau \omega_A = 5.56$ was selected so that the time average was over
one linear wave period of the initialized KAW, $T = 2 \pi / \omega_0$
with $\omega_0 = 1.13 \omega_A$.  For all three cases, the gyrotropic
structure is clearly organized by the parallel resonant velocity of
the initialized wave, marked with a grey line, with little structure
depending on $v_\perp$.\footnote{While $C_{E_\parallel}(\tau \neq 0)$
  is quantitatively similar if we use $g_p$, $h_p$, or $f_p$ in its
  calculation, the additional terms in $h_p$ and $f_p$ obscure the
  structure around $v_\parallel = v_{\rm res}$ in the distributions
  themselves.} Such structure was seen in the electrostatic
simulations of Landau damping described in \cite{Klein:2016} and
\cite{Howes:2017a}. To focus on this $v_\parallel$ dependence, we
calculate the reduced field-particle correlation, integrated over
$v_\perp$, which for notational simplicity, we write as
$C_{E_\parallel}(v_\parallel) = \int dv_\perp
C_{E_\parallel}(v_\parallel,v_\perp)$.

\begin{figure}
  \centerline{\includegraphics[width = 18cm]
    {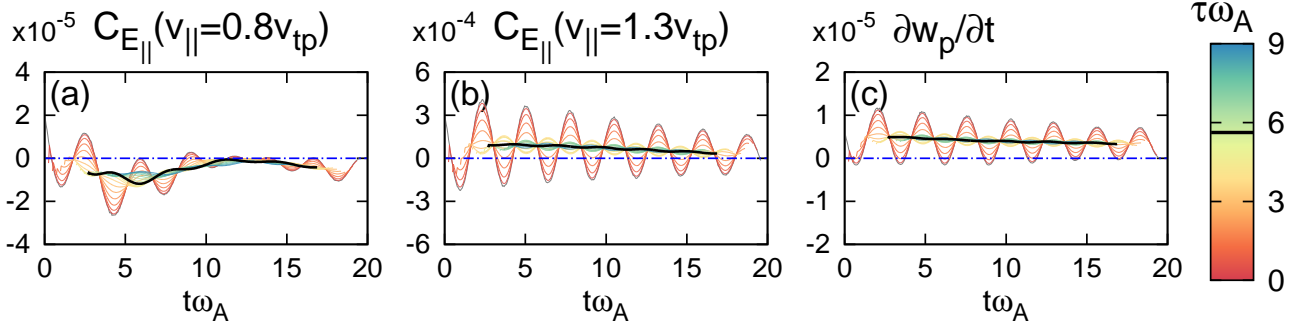}}%
  \caption{The reduced field-particle correlation
    $C_{E_\parallel}(v_\parallel)$ at two values of $v_\parallel$,
    panels (a) and (b), as well as the $v_\parallel$ integrated
    correlation $\partial w_p/\partial t$, panel (c), for a range of
    correlation intervals $\tau$ indicated by the colorbar. The
    correlation interval $\tau \omega_A =5.56$ selected for of
    Fig.~\ref{fig:gyro_KAW}(c) and Fig.~\ref{fig:2d_corr_KAW} is
    indicated with a black line.}
\label{fig:tau_KAW}
\end{figure}

 To illustrate the effects of the length of the correlation interval,
 in Fig.~\ref{fig:tau_KAW} we plot $C_{E_\parallel}(v_\parallel)$ for
 two values of $v_\parallel$ above and below the resonant velocity,
 $0.8 v_{tp}$ and $1.3 v_{tp}$, as well as the correlation integrated
 over $v_\parallel$, $\partial w_p(\V{r}_0,t)/\partial t = \int
 dv_\parallel C_{E_\parallel}(v_\parallel)$ where $w_p(\V{r}_0,t)$ is
 the ion spatial kinetic energy density at position $\V{r}_0$ and time
 $t$. We see that for an interval of exactly one linear wave period,
 the oscillatory component of the phase-space energy transfer is
 removed completely.  For correlation intervals that are not integral
 multiples of the wave period, the cancellation of the oscillatory
 component is not exact, but for correlation intervals longer than the
 wave period, $\tau>T$, the oscillatory component is significantly
 reduced, enabling the secular energy transfer associated with
 collisionless damping to be observed.  Note that the correlation
 $C_{E_\parallel}$ measures the rate of the change of phase space
 energy density for a particle species due to energy transfer with the
 fields; for sufficiently long correlation intervals, the net transfer
 in Fig.~\ref{fig:tau_KAW}(b) is positive showing that electric field
 is losing energy to the protons. As turbulence simulations will have
 a broadband spectrum of fluctuations with different periods, we will
 choose correlation intervals longer than the associated linear wave
 periods in an attempt to remove as much oscillatory energy transfer
 as possible.

\begin{figure}
  \centerline{\includegraphics[width = 12cm]
    {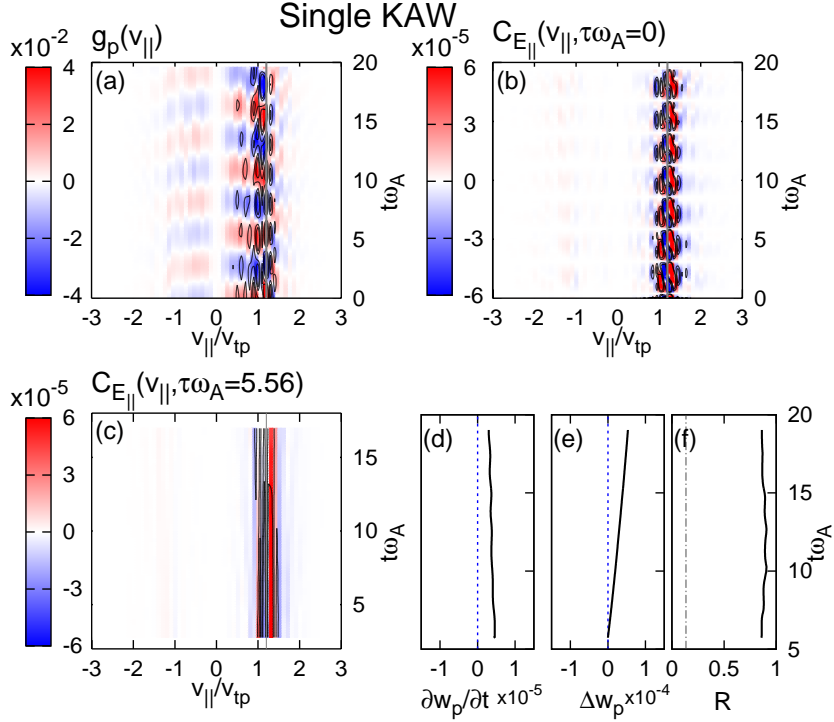}}%
  \caption{Timestack plots from the single KAW simulation, showing (a)
    the structure of the reduced complementary proton distribution
    function $g_p(v_\parallel)$, (b) the instantaneous field-particle
    correlation $C_{E_\parallel}(v_\parallel,\tau = 0)$, (c) the
    averaged field-particle correlations
    $C_{E_\parallel}(v_\parallel,\tau = 2 \pi /\omega_0)$, (d) the
    rate of change in the ion kinetic energy density $\partial
    w_p(t)/\partial t$, and (e) the net energy density transfer rate
    to the ions $\Delta w_p(t)$. The fraction of the energy
    transferred in the region around the resonant velocity of the KAW,
    $R$, is given in panel (f).}
\label{fig:2d_corr_KAW}
\end{figure}

The gyrotropic velocity-space plots in Fig.~\ref{fig:gyro_KAW} only
illustrate the energy transfer at a single point in time, but we are
interested in characterizing the entire time evolution.  Thus, we
integrate $g_p(v_\parallel,v_\perp,t)$ and
$C_{E_\parallel}(v_\parallel,v_\perp,t)$ over $v_\perp$ to obtain the
parallel reduced distribution function $g_p(v_\parallel,t)$ and
parallel reduced correlation $C_{E_\parallel}(v_\parallel,t)$; these
reduced values are then used to construct timestack plots that are
functions of only $v_\parallel$ and $t$, presented in
Fig.~\ref{fig:2d_corr_KAW}. As with the gyrotropic distributions in
Fig.~\ref{fig:gyro_KAW}, the variations as a function of $v_\parallel$
in Fig.~\ref{fig:2d_corr_KAW}, including (a) the reduced complementary
distribution function $\int dv_\perp v_\perp g_p$, as well as the
correlations (b) $C_{E_\parallel}(v_\parallel,\tau = 0)$ and (c)
$C_{E_\parallel}(v_\parallel,\tau = 2 \pi /\omega_0=5.56 /\omega_A)$,
are all organized about the resonant velocity of the initialized
KAW. However, in these timestack plots, we see the significant
oscillatory behavior in time in both the velocity distribution
function and the instantaneous phase-space energy transfer, while the
correlation averaged over one linear wave period reveals the secular,
resonant energy transfer. Note that the signature of energy gain above
the resonant velocity (red) and energy loss below the resonant
velocity (blue) in panel (c) corresponds to the flattening of the
distribution function found in quasilinear treatments of collisionless
damping \citep{Klein:2016,Howes:2017a}. This is the velocity-space
signature of the ion Landau damping of the kinetic \Alfven wave.

To track the total rate of energy transfer at this point in coordinate
space, we plot in Fig.~\ref{fig:2d_corr_KAW}(d) the velocity-space
integrated correlation $\partial w_p(\V{r}_0,t)/\partial t$, which
represents the total rate of energy transfer between the parallel
electric field and the ions at that position in space. In
Fig.~\ref{fig:2d_corr_KAW}(e), we plot the accumulated energy transfer
to the ions at position $\V{r}_0$, given by $\Delta w_p(\V{r}_0,t) =
\int_0^{t}dt' \partial w_p(\V{r}_0,t')/\partial t'$. These two
measures show that the physical mechanism of Landau damping achieves a
net transfer of energy to the ions over time at this position in the
simulation, as expected for a collisionlessly damped KAW.

To better quantify the resonant nature of the secular energy transfer,
we define the ratio
\begin{equation}
R\equiv
\frac{\int_{v_1}^{v_2}dv_\parallel|C_{E_\parallel}(v_\parallel)|}
{\int_{-4v_{tp}}^{4v_{tp}}dv_\parallel|C_{E_\parallel}(v_\parallel)|}
\label{eqn:R}
\end{equation}
where $v_1=0.65 v_{\rm res}$ and $v_2=1.35 v_{\rm res}$ and the
simulation domain extends from $v_\parallel = -4 v_{tp}$ to $4
v_{tp}$.  The values of $v_1$ and $v_2$ are selected so that $90\%$ of
the energy transferred is within the region between these two
velocities. The value of $R$ for the single KAW simulation is
presented in Fig.~\ref{fig:2d_corr_KAW}(f). We use this ratio to
assess how much of the energy transfer in turbulent simulations is due
to interactions with resonant particles. To help in the physical
interpretation of $R$, we estimate what fraction of the energy
transfer would be mediated by these particles if the energy transfer was
equally partitioned according to the equilibrium velocity
distribution.  That estimate, which is just the fraction of particles
within the resonant energy range from $v_1$ to $v_2$, given by
$\int_{v_1}^{v_2}dv_\parallel \exp [ -v_\parallel^2/v_{tp}^2]/
\int_{-4v_{tp}}^{4v_{tp}}dv_\parallel \exp [
  -v_\parallel^2/v_{tp}^2]$, has a value of $0.134$ (vertical grey
dot-dashed line), much smaller than the fraction computed from the
simulation, $R \simeq 0.9$, shown in
Fig.~\ref{fig:2d_corr_KAW}(f). Therefore, the resonant particles
dominate the energy transfer, as expected for the Landau damping
occurring in this system.

\section{Field-Particle Correlations in Strong Plasma Turbulence}
\label{sec:apply}

\subsection{Single-Point Field-Particle Correlations}
\label{ssec:singlePt}

\begin{figure}
  \centerline{\includegraphics[width = 18cm]
    {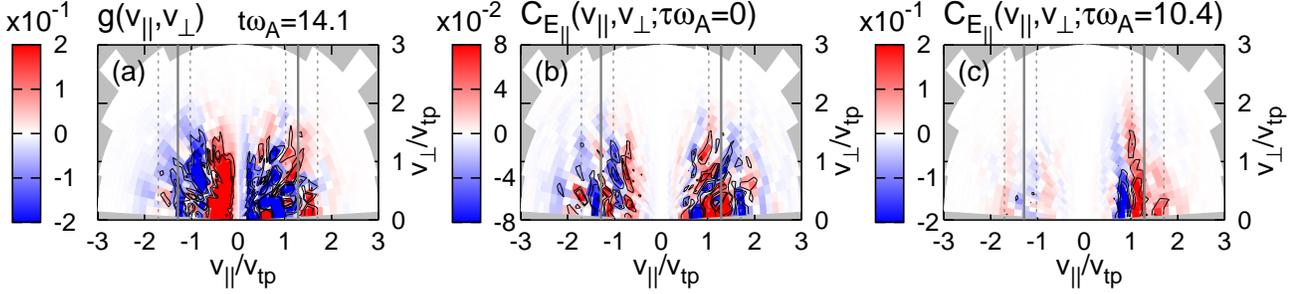}}%
  \caption{(a) The gyrotropic complementary distribution function
    $g_p(v_\parallel,v_\perp)$ at a single point in the $\beta_p=1.0$
    turbulent simulation, as well as the correlations (b)
    $C_{E_\parallel}(\tau=0)$ and (c) $C_{E_\parallel}(\tau \omega_A
    =10.4)$ at time $t\omega_A=14.1$. The resonant parallel velocity
    associated with the maximum proton damping rate is shown as a solid
    grey vertical line.}
\label{fig:gyro}
\end{figure}

With the single KAW results providing context for the interpretation
of field-particle correlation results, we next apply the
field-particle correlation technique to data from a single spatial
point $\V{r}_0=[x,y,z]=[0,10.2,0]\rho_p$ in the turbulent
$\beta_p=1.0$ simulation domain, where $[0,0,0]$ is the midpoint of
the simulation box. In panel (a) of Fig.~\ref{fig:gyro}, the
complementary gyrokinetic distribution function
$g_p(v_\parallel,v_\perp)$ is plotted at $\V{r}_0$ in the
$\beta_p=1.0$ run at a time sufficiently late in the run for the
turbulence to be fully developed, $t\omega_A= 14.1$. Solid grey lines
indicate the parallel resonant velocity for a KAW with the peak proton
damping rate, $v_{\rm res} = 1.282 v_{tp}$, and dashed lines indicate
the resonant velocities associated with KAWs having proton damping
rates equal to $1/e$ of the peak value, as identified in
Fig.~\ref{fig:linear}. We calculate the instantaneous phase-space
energy density transfer $C_{E_\parallel}(v_\parallel,v_\perp,\tau=0)$
in panel (b), and in panel (c), we calculate the correlation averaged
over an interval $\tau \omega_A =10.4$.

Unlike the case for a single KAW presented in
Fig.~\ref{fig:gyro_KAW}(a), Fig.~\ref{fig:gyro}(a) shows that the
structure of the complementary distribution function $g_p(v_\parallel,
v_\perp)$ for the strong turbulence simulation has large amplitude
fluctuations spread more broadly over velocity space, with the largest
amplitude fluctuations occurring at velocities $|v_\parallel| \ll
v_{tp}$.  Note also that, for the single KAW case in
Fig.~\ref{fig:gyro_KAW}(a), the fluctuations in $g_p(v_\parallel,
v_\perp)$ are almost entirely restricted to $v_\parallel > 0$; the
reason is because the wave is propagating in only one direction. In
the strong turbulence simulation shown in Fig.~\ref{fig:gyro}(a),
\Alfvenic fluctuations propagate in both directions, thereby leading
to significant fluctuations in $g_p(v_\parallel, v_\perp)$ at both
$v_\parallel > 0$ and $v_\parallel < 0$.

Taking the instantaneous correlation $C_{E_\parallel}$ with $\tau =0$
in Fig.~\ref{fig:gyro}(b), which corresponds to the rate of
instantaneous energy transfer between the parallel electric field and
the ions as a function of gyrotropic velocity space
$(v_\parallel,v_\perp)$, we see that the instantaneous energy transfer
is also broadly spread over a wide region of velocity space, with
significant structure as a function of $v_\perp$ and $v_\parallel$.
By taking the correlation over the interval $\tau \omega_A=10.4$,
equal to a correlation interval $\tau=1.67 T_0$ where $T_0$ is the
period of the largest scale \Alfven waves represented in the
simulation, we show in Fig.~\ref{fig:gyro}(c) that the energy transfer
is largely restricted to the region near the range of resonant
velocities for waves with the highest ion damping rates (within the
vertical dashed lines). In addition, the correlated energy transfer is
almost entirely a function of $v_\parallel$, with little significant
structure in $v_\perp$, as expected from kinetic theory for Landau
damping. It is remarkable that, even in a strong turbulence
simulation, the application of the field-particle correlation
technique obtains a velocity-space signature that is qualitatively
similar to the case for a single KAW, enabling a straightforward
interpretation of the results: the collisionless energy transfer
between the parallel electric field and the ions is dominantly a
resonant transfer associated with the Landau resonance, a key result
of this investigation.

\begin{figure}
  \centerline{\includegraphics[width = 18cm]
    {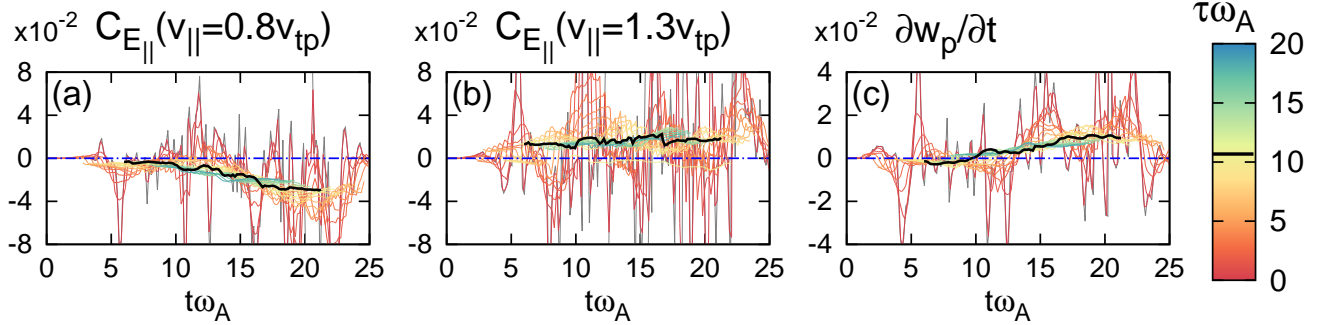}}%
  \caption{Reduced correlations $C_{E_\parallel}(v_\parallel)$ for (a)
    $v_\parallel=0.8 v_{tp}$ and (b) $v_\parallel=1.3 v_{tp}$, as well
    as (c) the velocity integrated $\partial w_p/\partial t$ for
    correlation intervals ranging from $0$ (grey) to
    $20/\omega_A$. Thick black lines indicate the correlation interval
    selected for Figs.~\ref{fig:gyro}, \ref{fig:vpar-b10}, and
    \ref{fig:Cprime}.}
\label{fig:dv}
\end{figure}

As with the single KAW case, the selection of the correlation interval
$\tau$ is crucial to separating the oscillatory and secular components
of the energy transfer. Choosing an appropriate interval is especially
challenging in the case of broadband turbulence as there is a spectrum
of frequencies associated with the secular energy transfer.  To study
the impact of the choice of particular values of $\tau$, we consider
the energy transfer captured by $C_{E_\parallel}(v_\parallel)$, at two
values of $v_\parallel$, as well as for the velocity integrated
correlation for a range of intervals, shown in Fig.~\ref{fig:dv}.

As discussed in \cite{Howes:2017a} and demonstrated in
Fig.~\ref{fig:tau_KAW} for the single KAW case, if the correlation
interval is longer than the wave period of a damping mode, the
oscillatory energy transfer will be largely averaged away, leaving
mostly the secular component.  As there is not a single wave period
for turbulent systems, we elect to average the field-particle
correlation over an interval much longer than the linear wave period
of the most strongly damped mode.  As shown in Fig.~\ref{fig:dv} with
black lines, the selected correlation interval of $\tau \omega_A =
10.4$ is sufficiently long to remove most of the fluctuations in the
energy transfer, leaving a nearly monotonic transfer of energy between
the fields and particles.

\begin{figure}
  \centerline{\includegraphics[width = 12cm]
    {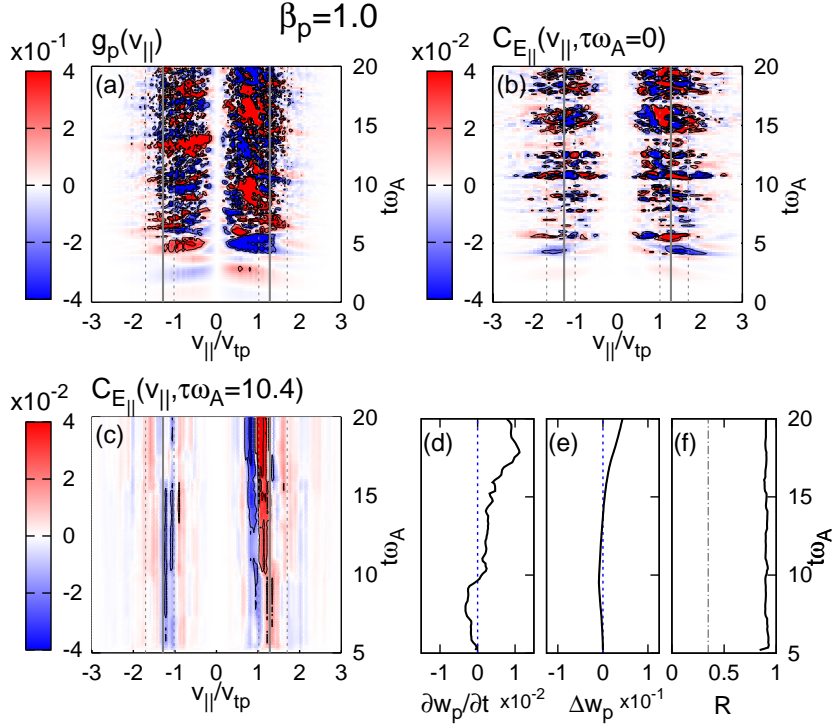}}%
  \caption{Timestack plots from the turbulent, $\beta_p = 1.0$
    simulation, using the same layout as presented in
    Fig.~\ref{fig:2d_corr_KAW}.}
\label{fig:vpar-b10}
\end{figure}

With a correlation interval $\tau \omega_A = 10.4$ selected to isolate
the secular energy transfer, we next present the timestack
distributions of $g_p(v_\parallel,t)$ and the associated reduced
field-particle correlations $C_{E_\parallel}(v_\parallel)$ from the
same position $\V{r}_0$ diagnosed in Fig.~\ref{fig:gyro}. The reduced
complementary proton velocity distribution $g_p(v_\parallel)$ is shown
in Fig.~\ref{fig:vpar-b10}(a). As with the gyrotropic plot of
$g_p(v_\parallel,v_\perp)$, there is no significant organization of
the structure of the distribution $g_p(v_\parallel)$ about the
preferred parallel resonant velocities of the system, but rather there
are large amplitude fluctuations at $|v_\parallel| \ll v_{tp}$. The
instantaneous rate of change of the phase-space energy density as a
function of $v_\parallel$, $C_{E_\parallel}(v_\parallel,\tau=0)$,
plotted in Fig.~\ref{fig:vpar-b10}(b), is broadly distributed about
the system's preferred resonant velocities. However, significant
oscillatory behavior in time is retained in the instantaneous energy
transfer.

The averaged correlation $C_{E_\parallel}(v_\parallel,\tau = 10.4)$ is
plotted in Fig.~\ref{fig:vpar-b10}(c), which shows clearly that the
net energy transfer is localized in the range of the resonant parallel
velocities. Again, this velocity-space signature clearly indicates
active Landau damping transferring energy to the ions via the parallel
electric field of the turbulent fluctuations. Tracking the energy
transfer rate at point $\V{r}_0$, we plot in panel (d) the
velocity-space integrated correlation $\partial
w_p(\V{r}_0,t)/\partial t$ and in panel (e) the accumulated energy
density transfer to the ions $\Delta w_p(\V{r}_0,t)$. These two
metrics show that a net ion energization over time occurs at this
position in the simulation.

While the resonant signature is not as clean as that seen in simpler
Vlasov-Poisson systems, or the single KAW simulation presented earlier
in this work, we can quantify the fraction of the energy transferred
by resonant particles using the ratio $R$, extended to include the
positive and negative resonant velocities. We set $v_1=\pm 0.65 v_{\rm
  res,lower}=0.66v_{tp}$ and $v_2=\pm 1.35 v_{\rm
  res,upper}=2.30v_{tp}$ and plot $R$ in
Fig.~\ref{fig:vpar-b10}(f). Over $92 \%$ of the net energy transferred
between fields and particles is mediated by the particles in this
resonance region. If the energy transfer was equally partitioned based
upon the particle density, we would expect only $35\%$ of the net
energy transfer to be carried by these particles (vertical grey
dot-dashed line).  Thus, this analysis shows clearly that a resonant
process is governing the net transfer of energy from fields to
particles.  The key result of this field-particle correlation analysis
is that this resonant process, Landau damping, is an effective
mechanism for the removal of energy from the turbulent fluctuations in
a strongly turbulent, kinetic plasma.

\subsection{Spatial Variation}
\label{ssec:spatial}

\begin{figure}
  \centerline{\includegraphics[width = 12cm]
    {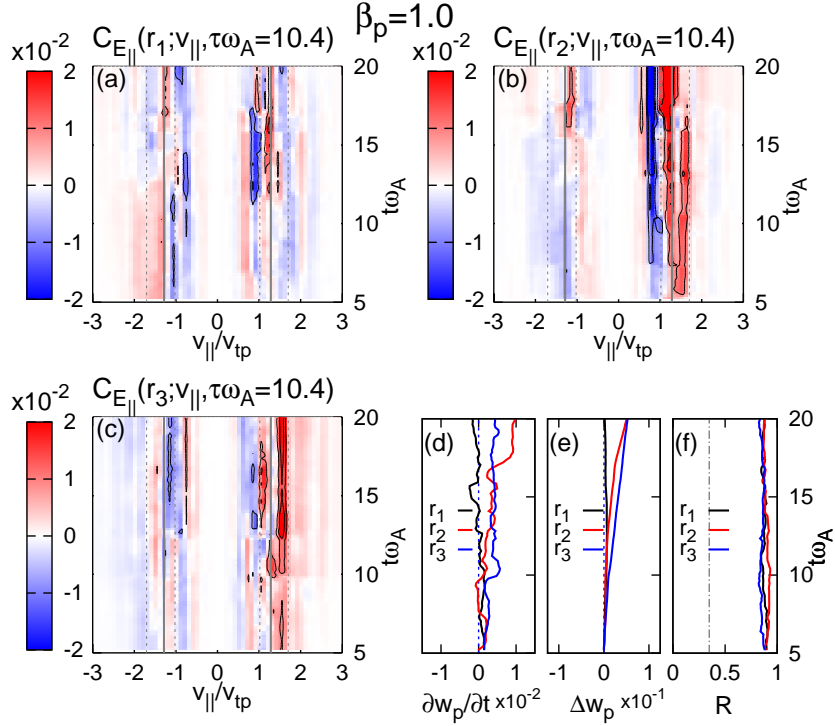}}%
  \caption{Reduced correlation $C_{E_\parallel}(v_\parallel,\tau
    \omega_A = 10.4)$ at three points $\V{r}_j$ (a-c) in the
    $\beta_p=1.0$ simulation, along with (d) the net energy transfer
    rate $\partial w_p(\V{r}_j,t)/\partial t$ and (e) the accumulated
    energy transfer $\Delta w_p(\V{r}_j,t)$. Panel (f) shows the
    fraction of the energy transferred in the region around the
    preferred resonant velocities, $R$.}
\label{fig:Cxpts}
\end{figure}

We next present timestack plots of the reduced correlation
$C_{E_\parallel}(v_\parallel,\tau \omega_A= 10.4)$ at three additional
distinct spatial points, $\V{r}_1=[6.6,6.6,0]\rho_p,
\ \V{r}_2=[0,12.6,0]\rho_p,$ and $\V{r}_3=[0,6.6,5.5/\epsilon]\rho_p$,
from the $\beta_p=1.0$ turbulent simulation in
Fig.~\ref{fig:Cxpts}(a)-(c). The $v_\parallel$ structure of the energy
transfer quantitatively differs between the three points but is
qualitatively organized by the resonant velocities for all three
cases. The net secular energy transfer, calculated by integrating over
$v_\parallel$, varies between these points, as shown in
Fig.~\ref{fig:Cxpts}(d) for the net energy density transfer rate
$\partial w_p(\V{r}_j,t)/\partial t$ and (e) for the accumulated
energy density transfer to the ions $\Delta w_p(\V{r}_j,t)$. The
spatial variation of the energy transfer to the ions is consistent
with previous findings that damping and heating in turbulent systems
is not spatially homogeneous but occurs intermittently in space
\citep{Wan:2012,Karimabadi:2013,TenBarge:2013a,Wu:2013a,Zhdankin:2013,Zhdankin:2015a}.
Further analysis at other points analyzed in the simulation domain
(not shown) demonstrates that, although the amplitude and sign of the
energy transfer differs from position to position, the energy transfer
between the parallel electric field and the ions is dominated by
resonant particles, all having values of $R \approx 0.9$, as shown in
Fig.~\ref{fig:Cxpts}(f). Therefore, this important result shows
definitively that Landau resonant collisionless energy transfer can
occur in a spatially non-uniform manner, in contrast to naive
expectations of Landau damping based on the plane-wave decomposition
usually used to derive linear Landau damping. Ongoing work using
field-particle correlations will determine whether Landau damping can
indeed be responsible for particle energization that is highly
intermittent in space, such as that occurring in the vicinity of
current sheets, as has been previously suggested
\citep{TenBarge:2013a,Howes:2015b,Howes:2016}.

\subsection{Variation with Plasma Beta}
\label{ssec:beta}

An important test of the application of the field-particle correlation
technique to strong plasma turbulence is the dependence of the results
on the plasma $\beta$.  At large scales $k_\perp \rho_p \ll 1$, the
value of the parallel \Alfven wave phase velocity normalized by the
\Alfven speed is simply unity, $\omega/(k_\parallel v_A)=1$.
Normalizing instead to the proton thermal velocity, this relation
becomes $\omega/(k_\parallel v_{tp})=v_A/v_{tp}=
\beta_p^{-1/2}$. Therefore, if the collisionless transfer of energy
between the electromagnetic fields and the ions is governed by a
resonant mechanism, the field-particle correlation technique will show
that the dominant regions of energy transfer in velocity space shift
accordingly as the plasma $\beta_p$ is changed.

\begin{figure}
  \centerline{\includegraphics[width = 12cm]
    {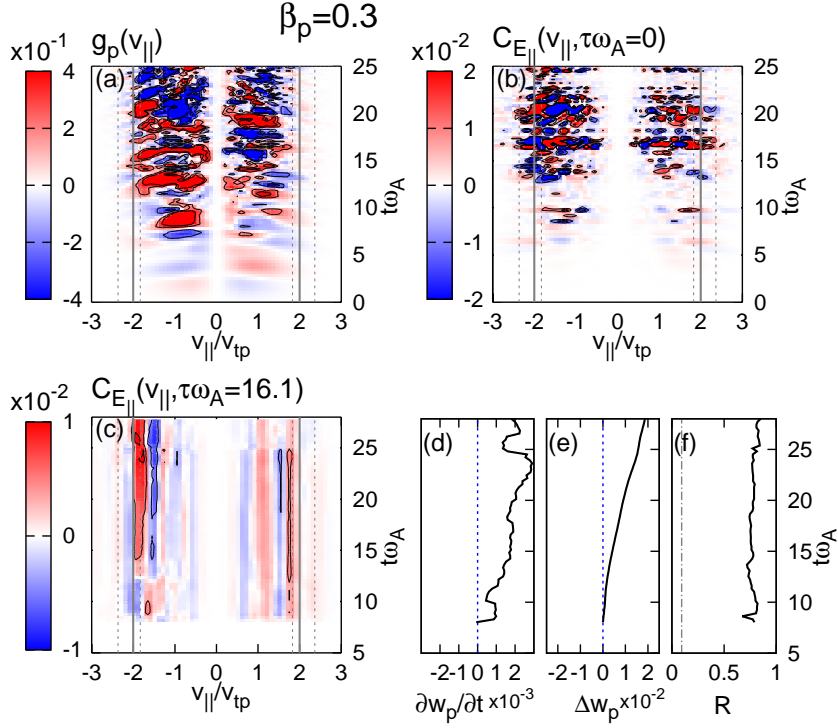}}%
  \caption{Timestack plots from the turbulent, $\beta_p = 0.3$
    simulation, using the same layout as presented in
    Fig.~\ref{fig:2d_corr_KAW}.}
\label{fig:vpar-b03}
\end{figure}

\begin{figure}
  \centerline{\includegraphics[width = 12cm]
    {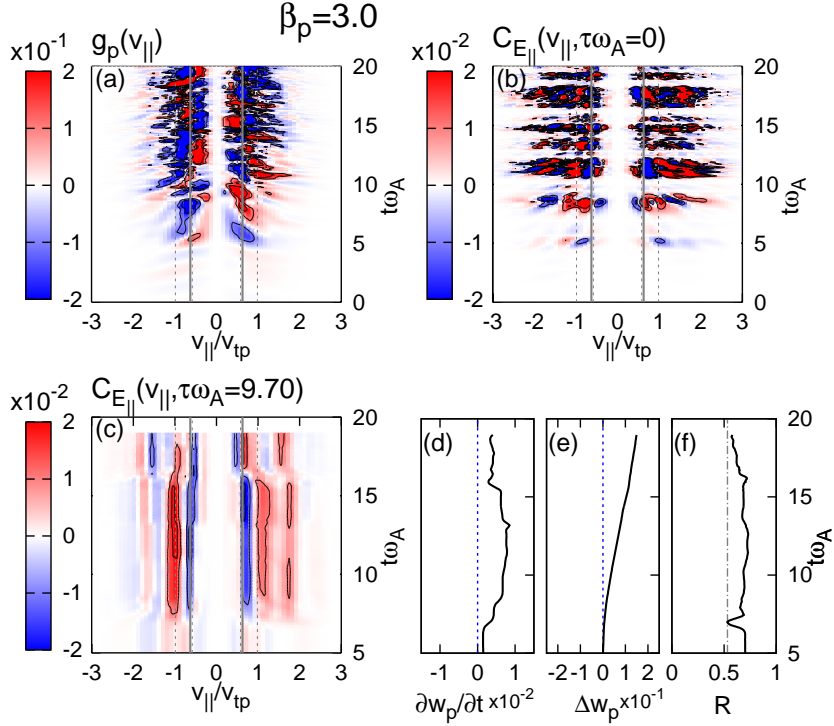}}%
  \caption{Timestack plots from the turbulent, $\beta_p = 3.0$
    simulation, using the same layout as presented in
    Fig.~\ref{fig:2d_corr_KAW}.}
\label{fig:vpar-b30}
\end{figure}

We present timestack plots of (a) the reduced complementary
distribution function $g_p(v_\parallel)=\int dv_\perp
g_p(v_\parallel,v_\perp)$, (b) the instantaneous phase-space energy
density transfer rate $C_{E_\parallel}(v_\parallel,\tau = 0)$, and (c)
the time-averaged correlation $C_{E_\parallel}(v_\parallel,\tau
\omega_A > 0 )$ for the $\beta_p = 0.3$ simulation in
Fig.~\ref{fig:vpar-b03} and for the $\beta_p = 3.0$ simulation in
Fig.~\ref{fig:vpar-b30}.  As expected from the scaling
$\omega/(k_\parallel v_{tp}) \propto \beta_p^{-1/2}$, illustrated in
Fig.~\ref{fig:linear}, the preferred resonant velocities are shifted
to higher $v_\parallel$ for lower $\beta_p$, and lower $v_\parallel$
for higher $\beta_p$, specifically $\omega/k_\parallel = 2.00 v_{tp}$
for $\beta_p = 0.3$ and $\omega/k_\parallel = 0.626 v_{tp}$ for
$\beta_p = 3.0$. We choose correlation intervals, $\tau \omega_A
= 16.1$ and $9.70$ for the $\beta_p = 0.3$ and $3.0$ simulations to
remove the oscillatory component of the energy transfer. In panels (d)
and (e), the net secular energy density transfer rate $\partial
w_p(\V{r}_j,t)/\partial t$ and the accummulated energy density transfer
$\Delta w_p(\V{r}_j,t)$ is plotted, showing a net transfer of energy
to the ions from the electric field.

The fraction of energy transferred by particles with parallel
velocities near the resonant velocities, $R$, is large for both
simulations. As with the $\beta_p=1.0$ turbulent simulation, we select
$|v_1|=0.65 v_{\rm res,lower}$ and $|v_2|=1.35 v_{\rm res,upper}$,
where $v_{\rm res,lower}$ and $v_{\rm res,upper}$ are the resonant
velocities associated with the KAW wavemodes having proton damping
rates equal to $1/e$ of the peak proton damping rate. This selection
results in $|v_1| = 1.188 (0.378) v_{tp}$ and $|v_2| = 3.193 (1.330)
v_{tp}$ for the $\beta_p = 0.3 (3.0)$ simulation, and yields $R=0.75
(0.7)$, shown in panel (f). If the energy transfer was equally
partitioned based upon velocity-space density, these particles would
only be responsible for $9 \%$ and $53 \%$ of the energy transfer
respectively.  Thus, for all examined values of $\beta_p$, the
phase-space energy transfer is largely consistent with the linear
predictions for resonant Landau damping. We note that we have
restricted this work to $\beta_p$ near unity to model typical 1
A.U. solar wind turbulence, restricting $v_{\parallel,\rm res}\sim
v_{tp}$.  Future work is underway to study the effects of significant
departures from $\beta_p = 1$ on the secular transfer of energy, as
both magnetically and thermally dominated plasmas are relevant in
different space and astrophysical contexts.

\subsection{Alternative Field-Particle Correlation $C_{E_\parallel}'$}
\label{ssec:Cprime}

\begin{figure}
  \centerline{\includegraphics[width = 12cm]
    {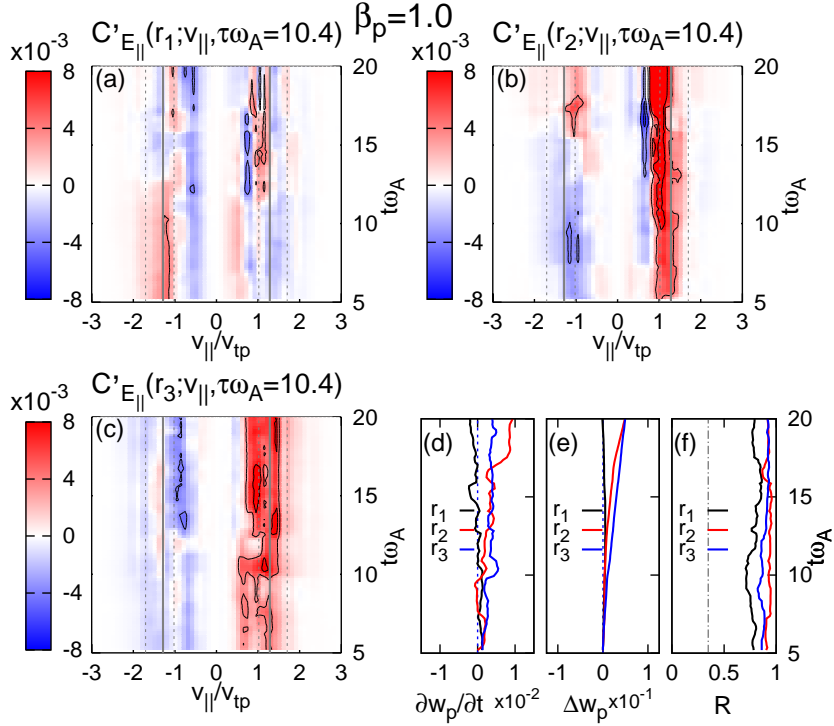}}%
  \caption{Reduced correlation $C_{E_\parallel}'(v_\parallel, \tau
    \omega_A= 10.4)$ and integrated quantities at the same three
    points in the $\beta_p=1.0$ simulation shown in
    Fig.~\ref{fig:Cxpts}.}
\label{fig:Cprime}
\end{figure}

As discussed in Sec.~\ref{sec:fpem}, limitations of spacecraft data
make velocity gradients $\partial f_s(\mathbf{v})/\partial
v_\parallel$ noisy and potentially unreliable. To alleviate this
problem, an alternative correlation, $C_{E_\parallel}'$, was
introduced in \eqref{eq:cprime_epar}. This alternative correlation is
calculated over the same correlation interval $\tau$ and at the same
three spatial points $\mathbf{r}_j$ used for Fig.~\ref{fig:Cxpts} and
is shown in Fig.~\ref{fig:Cprime}(a-c). As has been previously noted
for application of this technique to electrostatic systems, the
resonant signature---that is, the change in sign of the phase-space
energy transfer rate across a preferred velocity---is not present in
the structure of $C_{E_\parallel}'$. Nevertheless, a calculation of
the resonant fraction $R$ replacing $C_{E_\parallel}$ with
$C_{E_\parallel}'$, shows that the amplitude of the alternative
correlation remains significantly enhanced in the resonant particle
region. Comparison of the (d) net energy density transfer rate $\partial
w_p(\V{r}_j,t)/\partial t$ and (e) the accumulated energy density transfer to
the ions $\Delta w_p(\V{r}_j,t)$ for the standard correlation
$C_{E_\parallel}$ in Fig.~\ref{fig:Cxpts} and for the alternative
correlation $C_{E_\parallel}'$ in Fig.~\ref{fig:Cprime} shows that the
two forms of the correlation yield identical results. This must hold,
since the integrated quantities are related by an integration by parts
in velocity \citep{Howes:2017a}, both simply tracking the same
averaged $j_{p\parallel} E_\parallel$ energy transfer rate.  The
agreement serves as a check that both analysis methods are being
applied correctly.

\section{Conclusions and Future Work}
\label{sec:concl}

Here we have applied the field-particle correlation technique
\citep{Klein:2016,Howes:2017a} to explore the ion energization in
gyrokinetic simulations of strong plasma turbulence.  The results
definitively show that Landau damping persists as an effective
physical mechanism for ion energization in strong plasma turbulence,
contradicting recent suggestions that Landau damping may become
ineffective in the highly nonlinear environment of strong turbulence
\citep{Plunk:2013,Schekochihin:2016}. Furthermore, it is shown
directly that the ion energization resulting from the Landau damping
of turbulent electromagnetic fluctuations is spatially non-uniform, in
contrast to naive expectations that Landau damping leads to spatially
uniform energization, likely arising from the plane-wave decomposition
typically used to derive linear Landau damping. Further work using
field-particle correlations will address whether Landau damping can
effectively lead to the spatially intermittent plasma heating in the
vicinity of current sheets found in plasma turbulence simulations
\citep{Wan:2012,Karimabadi:2013,TenBarge:2013a,Wu:2013a,Zhdankin:2013}
and inferred from solar wind observations
\citep{Osman:2011,Osman:2012b,Perri:2012a,Wang:2013,Wu:2013a,Osman:2014b}.

Simulations with a wider range of plasma parameters than considered in
this work, especially more significant variations in $\beta_p$, will
be useful in further testing the applicability of this correlation to
a wide range of solar wind parameters.  This work only focuses on a
single class of dissipation mechanisms which satisfy the Landua
($n=0$) resonance. Future work will focus on characterizing the
velocity space structure of field-particle correlations due to other
damping mechanisms, including $n \neq 0$ cyclotron damping, stochastic
ion heating by low-frequency \Alfvenic turbulence, and energization
via magnetic reconnection.

This work demonstrates that field-particle correlations can be
usefully applied to data from single-point measurements of turbulent
space plasmas.  Application of this technique to current and proposed
missions, such as \textit{DSCOVR}, \textit{Magnetospheric
  Multiscale(MMS)}\citep{Burch:2016}, \textit{Solar Probe Plus}
\citep{Fox:2015} and \textit{Turbulence Heating ObserveR(THOR)}
\citep{Vaivads:2016} may enable the definitive identification of the
mechanisms which dissipate turbulence and heat the solar wind as it
expands through the heliosphere. The alternative correlation
$C_{E_\parallel}'$, which is easier to employ on noisy and lower
velocity-space resolution solar wind observations, is able to isolate
the regions in velocity space where the energy transfer is occurring.

The authors would like to thank Benjamin Chandran, Bill Dorland, and
Justin Kasper for helpful discussions during the execution of this
project. This work was supported by NASA grant HSR NNX16AM23G, NSF
grant CAREER AGS-1054061, NSF SHINE award AGS-1622306, and DOE
grant DE-SC0014599. This work used the Extreme Science and Engineering
Discovery Environment (XSEDE), which is supported by National Science
Foundation grant number ACI-1053575.

\bibliographystyle{jpp}

\bibliography{gkshort}

\begin{thebibliography}{74}
\expandafter\ifx\csname natexlab\endcsname\relax\def\natexlab#1{#1}\fi

\bibitem[{Abel} {\em et~al.\/}(2008){Abel}, {Barnes}, {Cowley}, {Dorland} \&
  {Schekochihin}]{Abel:2008}
{\sc {Abel}, I.~G., {Barnes}, M., {Cowley}, S.~C., {Dorland}, W. \&
  {Schekochihin}, A.~A.} 2008 {Linearized model Fokker-Planck collision
  operators for gyrokinetic simulations. I. Theory}. {\em Phys.~Plasmas\/} {\bf
  15}~(12), 122509.

\bibitem[{Barnes}(1966)]{Barnes:1966}
{\sc {Barnes}, A.} 1966 {Collisionless Damping of Hydromagnetic Waves}. {\em
  Phys.~Fluids\/} {\bf 9}, 1483--1495.

\bibitem[{Barnes} {\em et~al.\/}(2009){Barnes}, {Abel}, {Dorland}, {Ernst},
  {Hammett}, {Ricci}, {Rogers}, {Schekochihin} \& {Tatsuno}]{Barnes:2009}
{\sc {Barnes}, M., {Abel}, I.~G., {Dorland}, W., {Ernst}, D.~R., {Hammett},
  G.~W., {Ricci}, P., {Rogers}, B.~N., {Schekochihin}, A.~A. \& {Tatsuno}, T.}
  2009 {Linearized model Fokker-Planck collision operators for gyrokinetic
  simulations. II. Numerical implementation and tests}. {\em Phys.~Plasmas\/}
  {\bf 16}~(7), 072107.

\bibitem[{Belcher} \& {Davis}(1971)]{Belcher:1971}
{\sc {Belcher}, J.~W. \& {Davis}, Jr., L.} 1971 {Large-amplitude Alfv{\'e}n
  waves in the interplanetary medium, 2}. {\em J.~Geophys.~Res.\/} {\bf 76},
  3534.

\bibitem[{Bourouaine} \& {Chandran}(2013)]{Bourouaine:2013}
{\sc {Bourouaine}, S. \& {Chandran}, B.~D.~G.} 2013 {Observational Test of
  Stochastic Heating in Low-{$\beta$} Fast-solar-wind Streams}. {\em
  Astrophys.~J.\/} {\bf 774}, 96.

\bibitem[{Bourouaine} {\em et~al.\/}(2008){Bourouaine}, {Marsch} \&
  {Vocks}]{Bourouaine:2008}
{\sc {Bourouaine}, S., {Marsch}, E. \& {Vocks}, C.} 2008 {On the Efficiency of
  Nonresonant Ion Heating by Coronal Alfv{\'e}n Waves}. {\em
  Astrophys.~J.~Lett.\/} {\bf 684}, L119.

\bibitem[{Burch} {\em et~al.\/}(2016){Burch}, {Moore}, {Torbert} \&
  {Giles}]{Burch:2016}
{\sc {Burch}, J.~L., {Moore}, T.~E., {Torbert}, R.~B. \& {Giles}, B.~L.} 2016
  {Magnetospheric Multiscale Overview and Science Objectives}. {\em Space
  Sci.~Rev.\/} {\bf 199}, 5--21.

\bibitem[{Chandran}(2010)]{Chandran:2010b}
{\sc {Chandran}, B.~D.~G.} 2010 {Alfv{\'e}n-wave Turbulence and Perpendicular
  Ion Temperatures in Coronal Holes}. {\em Astrophys.~J.\/} {\bf 720},
  548--554.

\bibitem[{Chandran} {\em et~al.\/}(2010){Chandran}, {Li}, {Rogers}, {Quataert}
  \& {Germaschewski}]{Chandran:2010a}
{\sc {Chandran}, B.~D.~G., {Li}, B., {Rogers}, B.~N., {Quataert}, E. \&
  {Germaschewski}, K.} 2010 {Perpendicular Ion Heating by Low-frequency
  Alfv{\'e}n-wave Turbulence in the Solar Wind}. {\em Astrophys.~J.\/} {\bf
  720}, 503--515.

\bibitem[{Chen} {\em et~al.\/}(2013){Chen}, {Boldyrev}, {Xia} \&
  {Perez}]{Chen:2013a}
{\sc {Chen}, C.~H.~K., {Boldyrev}, S., {Xia}, Q. \& {Perez}, J.~C.} 2013
  {Nature of Subproton Scale Turbulence in the Solar Wind}. {\em
  Phys.~Rev.~Lett.\/} {\bf 110}~(22), 225002.

\bibitem[{Chen} {\em et~al.\/}(2001){Chen}, {Lin} \& {White}]{Chen:2001}
{\sc {Chen}, L., {Lin}, Z. \& {White}, R.} 2001 {On resonant heating below the
  cyclotron frequency}. {\em Phys.~Plasmas\/} {\bf 8}, 4713--4716.

\bibitem[{Coleman}(1968)]{Coleman:1968}
{\sc {Coleman}, Jr., P.~J.} 1968 {Turbulence, Viscosity, and Dissipation in the
  Solar-Wind Plasma}. {\em Astrophys.~J.\/} {\bf 153}, 371.

\bibitem[{Denskat} {\em et~al.\/}(1983){Denskat}, {Beinroth} \&
  {Neubauer}]{Denskat:1983}
{\sc {Denskat}, K.~U., {Beinroth}, H.~J. \& {Neubauer}, F.~M.} 1983
  {Interplanetary magnetic field power spectra with frequencies from 2.4 X 10
  to the -5th HZ to 470 HZ from HELIOS-observations during solar minimum
  conditions}. {\em Journal of Geophysics Zeitschrift Geophysik\/} {\bf 54},
  60--67.

\bibitem[{Dmitruk} {\em et~al.\/}(2004){Dmitruk}, {Matthaeus} \&
  {Seenu}]{Dmitruk:2004}
{\sc {Dmitruk}, P., {Matthaeus}, W.~H. \& {Seenu}, N.} 2004 {Test Particle
  Energization by Current Sheets and Nonuniform Fields in Magnetohydrodynamic
  Turbulence}. {\em Astrophys.~J.\/} {\bf 617}, 667--679.

\bibitem[{Fox} {\em et~al.\/}(2015){Fox}, {Velli}, {Bale}, {Decker},
  {Driesman}, {Howard}, {Kasper}, {Kinnison}, {Kusterer}, {Lario}, {Lockwood},
  {McComas}, {Raouafi} \& {Szabo}]{Fox:2015}
{\sc {Fox}, N.~J., {Velli}, M.~C., {Bale}, S.~D., {Decker}, R., {Driesman}, A.,
  {Howard}, R.~A., {Kasper}, J.~C., {Kinnison}, J., {Kusterer}, M., {Lario},
  D., {Lockwood}, M.~K., {McComas}, D.~J., {Raouafi}, N.~E. \& {Szabo}, A.}
  2015 {The Solar Probe Plus Mission: Humanity's First Visit to Our Star}. {\em
  Space Sci.~Rev.\/} .

\bibitem[{Frieman} \& {Chen}(1982)]{Frieman:1982}
{\sc {Frieman}, E.~A. \& {Chen}, L.} 1982 {Nonlinear gyrokinetic equations for
  low-frequency electromagnetic waves in general plasma equilibria}. {\em
  Phys.~Fluids\/} {\bf 25}, 502--508.

\bibitem[{Gary}(1999)]{Gary:1999}
{\sc {Gary}, S.~P.} 1999 {Collisionless dissipation wavenumber: Linear theory}.
  {\em J.~Geophys.~Res.\/} {\bf 104}, 6759--6762.

\bibitem[{Goldstein} {\em et~al.\/}(1994){Goldstein}, {Roberts} \&
  {Fitch}]{Goldstein:1994}
{\sc {Goldstein}, M.~L., {Roberts}, D.~A. \& {Fitch}, C.~A.} 1994 {Properties
  of the fluctuating magnetic helicity in the inertial and dissipation ranges
  of solar wind turbulence}. {\em J.~Geophys.~Res.\/} {\bf 99}, 11519--11538.

\bibitem[{Hollweg} \& {Isenberg}(2002)]{Hollweg:2002}
{\sc {Hollweg}, J.~V. \& {Isenberg}, P.~A.} 2002 {Generation of the fast solar
  wind: A review with emphasis on the resonant cyclotron interaction}. {\em
  J.~Geophys.~Res.\/} {\bf 107}, 1147.

\bibitem[{Howes}(2015)]{Howes:2015b}
{\sc {Howes}, G.~G.} 2015 {A dynamical model of plasma turbulence in the solar
  wind}. {\em Philosophical Transactions of the Royal Society of London Series
  A\/} {\bf 373}, 20140145--20140145.

\bibitem[{Howes}(2016)]{Howes:2016}
{\sc {Howes}, G.~G.} 2016 {The Dynamical Generation of Current Sheets in
  Astrophysical Plasma Turbulence}. {\em Astrophys.~J.~Lett.\/} {\bf 827}, L28.

\bibitem[{Howes}(2017)]{Howes:2017b}
{\sc {Howes}, G.~G} 2017 {Ronald C. Davidson Award 2016: A Prospectus on
  Kinetic Heliophysics}. {\em PoP, submitted\/} .

\bibitem[{Howes} {\em et~al.\/}(2012){Howes}, {Bale}, {Klein}, {Chen}, {Salem}
  \& {TenBarge}]{Howes:2012a}
{\sc {Howes}, G.~G., {Bale}, S.~D., {Klein}, K.~G., {Chen}, C.~H.~K., {Salem},
  C.~S. \& {TenBarge}, J.~M.} 2012 {The Slow-mode Nature of Compressible Wave
  Power in Solar Wind Turbulence}. {\em Astrophys.~J.~Lett.\/} {\bf 753}, L19.

\bibitem[{Howes} {\em et~al.\/}(2006){Howes}, {Cowley}, {Dorland}, {Hammett},
  {Quataert} \& {Schekochihin}]{Howes:2006}
{\sc {Howes}, G.~G., {Cowley}, S.~C., {Dorland}, W., {Hammett}, G.~W.,
  {Quataert}, E. \& {Schekochihin}, A.~A.} 2006 {Astrophysical Gyrokinetics:
  Basic Equations and Linear Theory}. {\em Astrophys.~J.\/} {\bf 651},
  590--614.

\bibitem[{Howes} {\em et~al.\/}(2008){Howes}, {Dorland}, {Cowley}, {Hammett},
  {Quataert}, {Schekochihin} \& {Tatsuno}]{Howes:2008a}
{\sc {Howes}, G.~G., {Dorland}, W., {Cowley}, S.~C., {Hammett}, G.~W.,
  {Quataert}, E., {Schekochihin}, A.~A. \& {Tatsuno}, T.} 2008 {Kinetic
  Simulations of Magnetized Turbulence in Astrophysical Plasmas}. {\em
  Phys.~Rev.~Lett.\/} {\bf 100}~(6), 065004.

\bibitem[Howes {\em et~al.\/}(2017)Howes, Klein \& Li]{Howes:2017a}
{\sc Howes, Gregory~G., Klein, Kristopher~G. \& Li, Tak~Chu} 2017 Diagnosing
  collisionless energy transfer using field–particle correlations:
  Vlasov–poisson plasmas. {\em J.~Plasma Phys.\/} {\bf 83}~(1).

\bibitem[{Howes} {\em et~al.\/}(2014){Howes}, {Klein} \&
  {TenBarge}]{Howes:2014a}
{\sc {Howes}, G.~G., {Klein}, K.~G. \& {TenBarge}, J.~M.} 2014 {Validity of the
  Taylor Hypothesis for Linear Kinetic Waves in the Weakly Collisional Solar
  Wind}. {\em Astrophys.~J.\/} {\bf 789}, 106.

\bibitem[{Howes} {\em et~al.\/}(2011){Howes}, {Tenbarge}, {Dorland},
  {Quataert}, {Schekochihin}, {Numata} \& {Tatsuno}]{Howes:2011a}
{\sc {Howes}, G.~G., {Tenbarge}, J.~M., {Dorland}, W., {Quataert}, E.,
  {Schekochihin}, A.~A., {Numata}, R. \& {Tatsuno}, T.} 2011 {Gyrokinetic
  Simulations of Solar Wind Turbulence from Ion to Electron Scales}. {\em
  Phys.~Rev.~Lett.\/} {\bf 107}~(3), 035004.

\bibitem[{Isenberg} \& {Hollweg}(1983)]{Isenberg:1983}
{\sc {Isenberg}, P.~A. \& {Hollweg}, J.~V.} 1983 {On the preferential
  acceleration and heating of solar wind heavy ions}. {\em J.~Geophys.~Res.\/}
  {\bf 88}, 3923--3935.

\bibitem[{Johnson} \& {Cheng}(2001)]{Johnson:2001}
{\sc {Johnson}, J.~R. \& {Cheng}, C.~Z.} 2001 {Stochastic ion heating at the
  magnetopause due to kinetic Alfv{\'e}n waves}. {\em Geophys.~Res.~Lett.\/}
  {\bf 28}, 4421--4424.

\bibitem[{Karimabadi} {\em et~al.\/}(2013){Karimabadi}, {Roytershteyn}, {Wan},
  {Matthaeus}, {Daughton}, {Wu}, {Shay}, {Loring}, {Borovsky}, {Leonardis},
  {Chapman} \& {Nakamura}]{Karimabadi:2013}
{\sc {Karimabadi}, H., {Roytershteyn}, V., {Wan}, M., {Matthaeus}, W.~H.,
  {Daughton}, W., {Wu}, P., {Shay}, M., {Loring}, B., {Borovsky}, J.,
  {Leonardis}, E., {Chapman}, S.~C. \& {Nakamura}, T.~K.~M.} 2013 {Coherent
  structures, intermittent turbulence, and dissipation in high-temperature
  plasmas}. {\em Physics of Plasmas\/} {\bf 20}~(1), 012303.

\bibitem[{Kiyani} {\em et~al.\/}(2013){Kiyani}, {Chapman}, {Sahraoui}, {Hnat},
  {Fauvarque} \& {Khotyaintsev}]{Kiyani:2013}
{\sc {Kiyani}, K.~H., {Chapman}, S.~C., {Sahraoui}, F., {Hnat}, B.,
  {Fauvarque}, O. \& {Khotyaintsev}, Y.~V.} 2013 {Enhanced Magnetic
  Compressibility and Isotropic Scale Invariance at Sub-ion Larmor Scales in
  Solar Wind Turbulence}. {\em Astrophys.~J.\/} {\bf 763}, 10.

\bibitem[{Klein}(2017)]{Klein:2017}
{\sc {Klein}, K.~G} 2017 {Characterizing Fluid and Kinetic Instabilities using
  Field-Particle Correlations on Single-Point Time Series}. {\em ArXiv
  e-prints\/} .

\bibitem[{Klein} \& {Chandran}(2016)]{Klein:2016a}
{\sc {Klein}, K.~G. \& {Chandran}, B.~D.~G.} 2016 {Evolution of The Proton
  Velocity Distribution due to Stochastic Heating in the Near-Sun Solar Wind}.
  {\em Astrophys.~J.\/} {\bf 820}, 47.

\bibitem[{Klein} \& {Howes}(2015)]{Klein:2015a}
{\sc {Klein}, K.~G. \& {Howes}, G.~G.} 2015 {Predicted impacts of proton
  temperature anisotropy on solar wind turbulence}. {\em Phys.~Plasmas\/} {\bf
  22}~(3), 032903.

\bibitem[Klein \& Howes(2016)]{Klein:2016}
{\sc Klein, K.~G. \& Howes, G.~G.} 2016 Measuring collisionless damping in
  heliospheric plasmas using field–particle correlations. {\em
  Astrophys.~J.~Lett.\/} {\bf 826}~(2), L30.

\bibitem[Landau(1946)]{Landau:1946}
{\sc Landau, L.~D.} 1946 {On the vibrations of the electronic plasma}. {\em J.
  Phys.(USSR)\/} {\bf 10}, 25--34, [Zh. Eksp. Teor. Fiz.16,574(1946)].

\bibitem[{Leamon} {\em et~al.\/}(1998){Leamon}, {Smith}, {Ness}, {Matthaeus} \&
  {Wong}]{Leamon:1998b}
{\sc {Leamon}, R.~J., {Smith}, C.~W., {Ness}, N.~F., {Matthaeus}, W.~H. \&
  {Wong}, H.~K.} 1998 {Observational constraints on the dynamics of the
  interplanetary magnetic field dissipation range}. {\em J.~Geophys.~Res.\/}
  {\bf 103}, 4775.

\bibitem[{Lehe} {\em et~al.\/}(2009){Lehe}, {Parrish} \& {Quataert}]{Lehe:2009}
{\sc {Lehe}, R., {Parrish}, I.~J. \& {Quataert}, E.} 2009 {The Heating of Test
  Particles in Numerical Simulations of Alfv{\'e}nic Turbulence}. {\em
  Astrophys.~J.\/} {\bf 707}, 404--419.

\bibitem[{Markovskii} \& {Vasquez}(2011)]{Markovskii:2011}
{\sc {Markovskii}, S.~A. \& {Vasquez}, B.~J.} 2011 {A Short-timescale Channel
  of Dissipation of the Strong Solar Wind Turbulence}. {\em Astrophys.~J.\/}
  {\bf 739}, 22.

\bibitem[{Matthaeus} \& {Velli}(2011)]{Matthaeus:2011}
{\sc {Matthaeus}, W.~H. \& {Velli}, M.} 2011 {Who Needs Turbulence?. A Review
  of Turbulence Effects in the Heliosphere and on the Fundamental Process of
  Reconnection}. {\em Space Sci.~Rev.\/} {\bf 160}, 145--168.

\bibitem[{McChesney} {\em et~al.\/}(1987){McChesney}, {Stern} \&
  {Bellan}]{McChesney:1987}
{\sc {McChesney}, J.~M., {Stern}, R.~A. \& {Bellan}, P.~M.} 1987 {Observation
  of fast stochastic ion heating by drift waves}. {\em Phys.~Rev.~Lett.\/} {\bf
  59}, 1436--1439.

\bibitem[{Narita} {\em et~al.\/}(2011){Narita}, {Gary}, {Saito}, {Glassmeier}
  \& {Motschmann}]{Narita:2011}
{\sc {Narita}, Y., {Gary}, S.~P., {Saito}, S., {Glassmeier}, K.-H. \&
  {Motschmann}, U.} 2011 {Dispersion relation analysis of solar wind
  turbulence}. {\em Geophys.~Res.~Lett.\/} {\bf 38}, 5101.

\bibitem[{Nielson} {\em et~al.\/}(2010){Nielson}, {Howes}, {Tatsuno}, {Numata}
  \& {Dorland}]{Nielson:2010}
{\sc {Nielson}, K.~D., {Howes}, G.~G., {Tatsuno}, T., {Numata}, R. \&
  {Dorland}, W.} 2010 {Numerical modeling of Large Plasma Device Alfv{\'e}n
  wave experiments using AstroGK}. {\em Phys.~Plasmas\/} {\bf 17}~(2),
  022105--022105.

\bibitem[{Numata} {\em et~al.\/}(2010){Numata}, {Howes}, {Tatsuno}, {Barnes} \&
  {Dorland}]{Numata:2010}
{\sc {Numata}, R., {Howes}, G.~G., {Tatsuno}, T., {Barnes}, M. \& {Dorland},
  W.} 2010 {AstroGK: Astrophysical gyrokinetics code}. {\em J.~Comp.~Phys.\/}
  {\bf 229}, 9347--9372.

\bibitem[{Numata} \& {Loureiro}(2015)]{Numata:2015}
{\sc {Numata}, R. \& {Loureiro}, N.~F.} 2015 {Ion and electron heating during
  magnetic reconnection in weakly collisional plasmas}. {\em J.~Plasma Phys.\/}
  {\bf 81}, 30201.

\bibitem[{Osman} {\em et~al.\/}(2014{\natexlab{{\em a\/}}}){Osman}, {Kiyani},
  {Chapman} \& {Hnat}]{Osman:2014b}
{\sc {Osman}, K.~T., {Kiyani}, K.~H., {Chapman}, S.~C. \& {Hnat}, B.}
  2014{\natexlab{{\em a\/}}} {Anisotropic Intermittency of Magnetohydrodynamic
  Turbulence}. {\em Astrophys.~J.~Lett.\/} {\bf 783}, L27.

\bibitem[{Osman} {\em et~al.\/}(2014{\natexlab{{\em b\/}}}){Osman},
  {Matthaeus}, {Gosling}, {Greco}, {Servidio}, {Hnat}, {Chapman} \&
  {Phan}]{Osman:2014a}
{\sc {Osman}, K.~T., {Matthaeus}, W.~H., {Gosling}, J.~T., {Greco}, A.,
  {Servidio}, S., {Hnat}, B., {Chapman}, S.~C. \& {Phan}, T.~D.}
  2014{\natexlab{{\em b\/}}} {Magnetic Reconnection and Intermittent Turbulence
  in the Solar Wind}. {\em Physical Review Letters\/} {\bf 112}~(21), 215002.

\bibitem[{Osman} {\em et~al.\/}(2011){Osman}, {Matthaeus}, {Greco} \&
  {Servidio}]{Osman:2011}
{\sc {Osman}, K.~T., {Matthaeus}, W.~H., {Greco}, A. \& {Servidio}, S.} 2011
  {Evidence for Inhomogeneous Heating in the Solar Wind}. {\em
  Astrophys.~J.~Lett.\/} {\bf 727}, L11.

\bibitem[{Osman} {\em et~al.\/}(2012){Osman}, {Matthaeus}, {Wan} \&
  {Rappazzo}]{Osman:2012b}
{\sc {Osman}, K.~T., {Matthaeus}, W.~H., {Wan}, M. \& {Rappazzo}, A.~F.} 2012
  {Intermittency and Local Heating in the Solar Wind}. {\em Physical Review
  Letters\/} {\bf 108}~(26), 261102.

\bibitem[{Perri} {\em et~al.\/}(2012){Perri}, {Goldstein}, {Dorelli} \&
  {Sahraoui}]{Perri:2012a}
{\sc {Perri}, S., {Goldstein}, M.~L., {Dorelli}, J.~C. \& {Sahraoui}, F.} 2012
  {Detection of Small-Scale Structures in the Dissipation Regime of Solar-Wind
  Turbulence}. {\em Phys.~Rev.~Lett.\/} {\bf 109}~(19), 191101.

\bibitem[{Plunk}(2013)]{Plunk:2013}
{\sc {Plunk}, G.~G.} 2013 {Landau damping in a turbulent setting}. {\em
  Phys.~Plasmas\/} {\bf 20}~(3), 032304.

\bibitem[{Podesta} \& {TenBarge}(2012)]{Podesta:2012}
{\sc {Podesta}, J.~J. \& {TenBarge}, J.~M.} 2012 {Scale dependence of the
  variance anisotropy near the proton gyroradius scale: Additional evidence for
  kinetic Alfv{\'e}n waves in the solar wind at 1 AU}. {\em J.~Geophys.~Res.\/}
  {\bf 117}, A10106.

\bibitem[{Quataert}(1998)]{Quataert:1998}
{\sc {Quataert}, E.} 1998 {Particle Heating by Alfvenic Turbulence in Hot
  Accretion Flows}. {\em Astrophys.~J.\/} {\bf 500}, 978.

\bibitem[{Roberts} {\em et~al.\/}(2015){Roberts}, {Li} \&
  {Jeska}]{Roberts:2015b}
{\sc {Roberts}, O.~W., {Li}, X. \& {Jeska}, L.} 2015 {A Statistical Study of
  the Solar Wind Turbulence at Ion Kinetic Scales Using the k-filtering
  Technique and Cluster Data}. {\em Astrophys.~J.\/} {\bf 802}, 2.

\bibitem[{Roberts} {\em et~al.\/}(2013){Roberts}, {Li} \& {Li}]{Roberts:2013}
{\sc {Roberts}, O.~W., {Li}, X. \& {Li}, B.} 2013 {Kinetic Plasma Turbulence in
  the Fast Solar Wind Measured by Cluster}. {\em Astrophys.~J.\/} {\bf 769},
  58.

\bibitem[{Sahraoui} {\em et~al.\/}(2010){Sahraoui}, {Belmont}, {Goldstein} \&
  {Rezeau}]{Sahraoui:2010b}
{\sc {Sahraoui}, F., {Belmont}, G., {Goldstein}, M.~L. \& {Rezeau}, L.} 2010
  {Limitations of multispacecraft data techniques in measuring wave number
  spectra of space plasma turbulence}. {\em J.~Geophys.~Res.\/} {\bf 115},
  4206.

\bibitem[{Salem} {\em et~al.\/}(2012){Salem}, {Howes}, {Sundkvist}, {Bale},
  {Chaston}, {Chen} \& {Mozer}]{Salem:2012}
{\sc {Salem}, C.~S., {Howes}, G.~G., {Sundkvist}, D., {Bale}, S.~D., {Chaston},
  C.~C., {Chen}, C.~H.~K. \& {Mozer}, F.~S.} 2012 {Identification of Kinetic
  Alfv{\'e}n Wave Turbulence in the Solar Wind}. {\em Astrophys.~J.~Lett.\/}
  {\bf 745}, L9.

\bibitem[{Schekochihin} {\em et~al.\/}(2009){Schekochihin}, {Cowley},
  {Dorland}, {Hammett}, {Howes}, {Quataert} \& {Tatsuno}]{Schekochihin:2009}
{\sc {Schekochihin}, A.~A., {Cowley}, S.~C., {Dorland}, W., {Hammett}, G.~W.,
  {Howes}, G.~G., {Quataert}, E. \& {Tatsuno}, T.} 2009 {Astrophysical
  Gyrokinetics: Kinetic and Fluid Turbulent Cascades in Magnetized Weakly
  Collisional Plasmas}. {\em Astrophys.~J.~Supp.\/} {\bf 182}, 310--377.

\bibitem[Schekochihin {\em et~al.\/}(2016)Schekochihin, Parker, Highcock,
  Dellar, Dorland \& Hammett]{Schekochihin:2016}
{\sc Schekochihin, A.~A., Parker, J.~T., Highcock, E.~G., Dellar, P.~J.,
  Dorland, W. \& Hammett, G.~W.} 2016 Phase mixing versus nonlinear advection
  in drift-kinetic plasma turbulence. {\em J.~Plasma Phys.\/} {\bf 82},
  905820212 (47 pages).

\bibitem[{Servidio} {\em et~al.\/}(2011){Servidio}, {Greco}, {Matthaeus},
  {Osman} \& {Dmitruk}]{Servidio:2011a}
{\sc {Servidio}, S., {Greco}, A., {Matthaeus}, W.~H., {Osman}, K.~T. \&
  {Dmitruk}, P.} 2011 {Statistical association of discontinuities and
  reconnection in magnetohydrodynamic turbulence}. {\em Journal of Geophysical
  Research (Space Physics)\/} {\bf 116}, A09102.

\bibitem[{Stix}(1992)]{Stix:1992}
{\sc {Stix}, T.~H.} 1992 {\em {Waves in plasmas}\/}. American Institute of
  Physics.

\bibitem[{TenBarge} {\em et~al.\/}(2014{\natexlab{{\em a\/}}}){TenBarge},
  {Daughton}, {Karimabadi}, {Howes} \& {Dorland}]{TenBarge:2014b}
{\sc {TenBarge}, J.~M., {Daughton}, W., {Karimabadi}, H., {Howes}, G.~G. \&
  {Dorland}, W.} 2014{\natexlab{{\em a\/}}} {Collisionless reconnection in the
  large guide field regime: Gyrokinetic versus particle-in-cell simulations}.
  {\em Phys.~Plasmas\/} {\bf 21}~(2), 020708.

\bibitem[{TenBarge} \& {Howes}(2012)]{TenBarge:2012a}
{\sc {TenBarge}, J.~M. \& {Howes}, G.~G.} 2012 {Evidence of critical balance in
  kinetic Alfv{\'e}n wave turbulence simulations}. {\em Phys.~Plasmas\/} {\bf
  19}~(5), 055901.

\bibitem[{TenBarge} \& {Howes}(2013)]{TenBarge:2013a}
{\sc {TenBarge}, J.~M. \& {Howes}, G.~G.} 2013 {Current Sheets and
  Collisionless Damping in Kinetic Plasma Turbulence}. {\em
  Astrophys.~J.~Lett.\/} {\bf 771}, L27.

\bibitem[{TenBarge} {\em et~al.\/}(2013){TenBarge}, {Howes} \&
  {Dorland}]{TenBarge:2013c}
{\sc {TenBarge}, J.~M., {Howes}, G.~G. \& {Dorland}, W.} 2013 {Collisionless
  Damping at Electron Scales in Solar Wind Turbulence}. {\em Astrophys.~J.\/}
  {\bf 774}, 139.

\bibitem[{TenBarge} {\em et~al.\/}(2014{\natexlab{{\em b\/}}}){TenBarge},
  {Howes}, {Dorland} \& {Hammett}]{TenBarge:2014}
{\sc {TenBarge}, J.~M., {Howes}, G.~G., {Dorland}, W. \& {Hammett}, G.~W.}
  2014{\natexlab{{\em b\/}}} {An oscillating Langevin antenna for driving
  plasma turbulence simulations}. {\em Computer Physics Communications\/} {\bf
  185}, 578--589.

\bibitem[{TenBarge} {\em et~al.\/}(2012){TenBarge}, {Podesta}, {Klein} \&
  {Howes}]{TenBarge:2012b}
{\sc {TenBarge}, J.~M., {Podesta}, J.~J., {Klein}, K.~G. \& {Howes}, G.~G.}
  2012 {Interpreting Magnetic Variance Anisotropy Measurements in the Solar
  Wind}. {\em Astrophys.~J.\/} {\bf 753}, 107.

\bibitem[{Vaivads} {\em et~al.\/}(2016){Vaivads}, {Retin{\`o}}, {Soucek},
  {Khotyaintsev}, {Valentini}, {Escoubet}, {Alexandrova}, {Andr{\'e}}, {Bale},
  {Balikhin}, {Burgess}, {Camporeale}, {Caprioli}, {Chen}, {Clacey}, {Cully},
  {de Keyser}, {Eastwood}, {Fazakerley}, {Eriksson}, {Goldstein}, {Graham},
  {Haaland}, {Hoshino}, {Ji}, {Karimabadi}, {Kucharek}, {Lavraud}, {Marcucci},
  {Matthaeus}, {Moore}, {Nakamura}, {Narita}, {Nemecek}, {Norgren},
  {Opgenoorth}, {Palmroth}, {Perrone}, {Pin{\c c}on}, {Rathsman}, {Rothkaehl},
  {Sahraoui}, {Servidio}, {Sorriso-Valvo}, {Vainio}, {V{\"o}r{\"o}s} \&
  {Wimmer-Schweingruber}]{Vaivads:2016}
{\sc {Vaivads}, A., {Retin{\`o}}, A., {Soucek}, J., {Khotyaintsev}, Y.~V.,
  {Valentini}, F., {Escoubet}, C.~P., {Alexandrova}, O., {Andr{\'e}}, M.,
  {Bale}, S.~D., {Balikhin}, M., {Burgess}, D., {Camporeale}, E., {Caprioli},
  D., {Chen}, C.~H.~K., {Clacey}, E., {Cully}, C.~M., {de Keyser}, J.,
  {Eastwood}, J.~P., {Fazakerley}, A.~N., {Eriksson}, S., {Goldstein}, M.~L.,
  {Graham}, D.~B., {Haaland}, S., {Hoshino}, M., {Ji}, H., {Karimabadi}, H.,
  {Kucharek}, H., {Lavraud}, B., {Marcucci}, F., {Matthaeus}, W.~H., {Moore},
  T.~E., {Nakamura}, R., {Narita}, Y., {Nemecek}, Z., {Norgren}, C.,
  {Opgenoorth}, H., {Palmroth}, M., {Perrone}, D., {Pin{\c c}on}, J.-L.,
  {Rathsman}, P., {Rothkaehl}, H., {Sahraoui}, F., {Servidio}, S.,
  {Sorriso-Valvo}, L., {Vainio}, R., {V{\"o}r{\"o}s}, Z. \&
  {Wimmer-Schweingruber}, R.~F.} 2016 {Turbulence Heating ObserveR - satellite
  mission proposal}. {\em J.~Plasma Phys.\/} {\bf 82}~(5), 905820501.

\bibitem[{Wan} {\em et~al.\/}(2012){Wan}, {Matthaeus}, {Karimabadi},
  {Roytershteyn}, {Shay}, {Wu}, {Daughton}, {Loring} \& {Chapman}]{Wan:2012}
{\sc {Wan}, M., {Matthaeus}, W.~H., {Karimabadi}, H., {Roytershteyn}, V.,
  {Shay}, M., {Wu}, P., {Daughton}, W., {Loring}, B. \& {Chapman}, S.~C.} 2012
  {Intermittent Dissipation at Kinetic Scales in Collisionless Plasma
  Turbulence}. {\em Phys.~Rev.~Lett.\/} {\bf 109}~(19), 195001.

\bibitem[{Wang} {\em et~al.\/}(2013){Wang}, {Tu}, {He}, {Marsch} \&
  {Wang}]{Wang:2013}
{\sc {Wang}, X., {Tu}, C., {He}, J., {Marsch}, E. \& {Wang}, L.} 2013 {On
  Intermittent Turbulence Heating of the Solar Wind: Differences between
  Tangential and Rotational Discontinuities}. {\em Astrophys.~J.~Lett.\/} {\bf
  772}, L14.

\bibitem[{Wu} {\em et~al.\/}(2013){Wu}, {Perri}, {Osman}, {Wan}, {Matthaeus},
  {Shay}, {Goldstein}, {Karimabadi} \& {Chapman}]{Wu:2013a}
{\sc {Wu}, P., {Perri}, S., {Osman}, K., {Wan}, M., {Matthaeus}, W.~H., {Shay},
  M.~A., {Goldstein}, M.~L., {Karimabadi}, H. \& {Chapman}, S.} 2013
  {Intermittent Heating in Solar Wind and Kinetic Simulations}. {\em
  Astrophys.~J.~Lett.\/} {\bf 763}, L30.

\bibitem[{Zhdankin} {\em et~al.\/}(2015){Zhdankin}, {Uzdensky} \&
  {Boldyrev}]{Zhdankin:2015a}
{\sc {Zhdankin}, V., {Uzdensky}, D.~A. \& {Boldyrev}, S.} 2015 {Temporal
  Intermittency of Energy Dissipation in Magnetohydrodynamic Turbulence}. {\em
  Physical Review Letters\/} {\bf 114}~(6), 065002.

\bibitem[{Zhdankin} {\em et~al.\/}(2013){Zhdankin}, {Uzdensky}, {Perez} \&
  {Boldyrev}]{Zhdankin:2013}
{\sc {Zhdankin}, V., {Uzdensky}, D.~A., {Perez}, J.~C. \& {Boldyrev}, S.} 2013
  {Statistical Analysis of Current Sheets in Three-dimensional
  Magnetohydrodynamic Turbulence}. {\em Astrophys.~J.\/} {\bf 771}, 124.

\end{thebibliography}

\end{document}